\documentclass[A4paper,twoside,journall]{IEEEtran}
\usepackage{amsmath,amsfonts}
\usepackage{algorithmic}
\usepackage{algorithm}
\usepackage{array}
\usepackage[caption=false,font=normalsize,labelfont=sf,textfont=sf]{subfig}
\usepackage{textcomp}
\usepackage{stfloats}
\usepackage{url}
\usepackage{verbatim}
\usepackage{graphicx}
\usepackage{enumitem}
\usepackage{cite}
\usepackage[]{mdframed}
\usepackage[utf8]{inputenc}
\usepackage{booktabs}
\usepackage{arydshln}
\usepackage{multirow}
\usepackage{svg}
\usepackage{amssymb}
\usepackage{pifont}
\usepackage[dvipsnames]{xcolor}
\usepackage{threeparttable}
\usepackage[table,xcdraw]{xcolor} 
\definecolor{deeppink}{RGB}{255,20,147}
\definecolor{darkpurple}{RGB}{128,0,128}
\definecolor{wine}{RGB}{128,0,32}
\def\cnt{\stepcounter{enumi}\arabic{enumi}}

     \usepackage[compact]{titlesec}
    \titlespacing{\section}{0pt}{2ex}{1ex}
    \titlespacing{\subsection}{0pt}{1ex}{0ex}
    \titlespacing{\subsubsection}{0pt}{0.5ex}{0ex}

\usepackage[colorlinks=true,linkcolor=blue]{hyperref}%

\usepackage{tikz}
\newcommand*\circled[1]{\tikz[baseline=(char.base)]{
            \node[shape=circle,draw,inner sep=0.8pt] (char) {#1};}}
            
\DeclareMathOperator*{\argmax}{\mathtt{argmax}}

\newcommand{\cb}[1]{\colorbox{blue!35}{#1}} % Best (Xanh đậm)
\newcommand{\cs}[1]{\colorbox{blue!15}{#1}} % Second best (Xanh nhạt)
\newcommand{\cw}[1]{\colorbox{red!35}{#1}}  % Worst (Đỏ đậm)
\newcommand{\csw}[1]{\colorbox{red!15}{#1}} % Second worst (Đỏ nhạt)
\newcommand{\FISHER}{\mathtt{FISHER}}

\makeatletter
\g@addto@macro\normalsize{%
 \setlength\abovedisplayskip{3pt}
 \setlength\belowdisplayskip{3pt}
 \setlength\abovedisplayshortskip{3pt}
 \setlength\belowdisplayshortskip{3pt}
}

\newcommand{\ph}[1]{\textcolor{black}{#1}}
\begin{document}

\title{FISHER: Gradient-Decoupled Hierarchical Multi-Task Learning for Fine-Grained Aquatic Species Recognition}

\author{{Phuc H. Nguyen$^*$, Ba Hung Ngo$^*$, Mai Phuong Tran, Cuong D. Do and Van-Dinh Nguyen}

\thanks{$^{*}$Both authors contributed equally to this work.}
\thanks{P. H. Nguyen, M. P. Tran and C. D. Do are with Smart Green Transformation Center and College of Engineering and Computer Science, VinUniversity, Vietnam (e-mail: cuong.dd@vinuni.edu.vn). B. H. Ngo is with  Chonnam National University, South Korea, and also with VinUniversity, Vietnam  (e-mail: ngohung@jnu.ac.kr). V.-D. Nguyen (corresponding author) is with Trinity College Dublin,  Ireland  (e-mail: dinh.nguyen@tcd.ie). }

% This work is sponsored by VinUniversity under VUNI.INNO.AY24-25.05.

\vspace{-5pt}
}
\markboth{Submitted to IEEE Transactions on Circuits and Systems for Video Technology}
{Nguyen \MakeLowercase{\textit{et al.}}: Gradient-Decoupled Hierarchical Multi-Task Learning for Fine-Grained Aquatic Species Recognition}

\maketitle

\begin{abstract} 
Fine-grained recognition of aquatic species is challenging due to subtle morphological differences and long-tailed distributions, where ultra-rare species are underrepresented. A natural solution is to jointly model segmentation, morphological traits, and species classification within a multi-task learning (MTL) framework. However, existing MTL methods suffer from negative transfer caused by gradient conflicts between low-level dense tasks and high-level classification objectives, degrading fine-grained representations. To address this limitation, we identify gradient interference across hierarchical tasks as a fundamental bottleneck and propose $\FISHER$, a gradient-decoupled hierarchical multi-task learning framework. $\FISHER$ aligns optimization with the biological hierarchy of aquatic species by enforcing a unidirectional information flow from segmentation to trait prediction and finally to species classification, while explicitly decoupling gradients across task boundaries. This design prevents high-level objectives from corrupting low-level morphological representations, effectively mitigating negative transfer while preserving the benefits of shared supervision. Furthermore, we introduce a prototype-based segmentation head with orthogonality regularization to encourage disentangled anatomical representations, and employ homoscedastic uncertainty weighting to dynamically balance task contributions during training. Our analysis shows that robust trait representations serve as a critical bridge for transferring knowledge to ultra-rare species. \ph{Extensive experiments on the Fish-Vista benchmark demonstrate that $\FISHER$ achieves $97.7\%$ mAP for trait identification on unseen species and improves ultra-rare species classification accuracy by $13.4\%$ over strong baselines, highlighting the effectiveness of gradient-decoupled hierarchical learning for long-tailed biodiversity recognition.}

\end{abstract}


\section{Introduction}
\IEEEPARstart{F}{ine-grained} visual recognition in the aquatic domain is an important yet challenging problem in computer vision \cite{Fishvista, AutoFish, 11373236, FishNet, saleh2020deepfish, garcia2022deepfish}. In fish species recognition, inter-class differences are often subtle, involving fine morphological details such as the caudal fin, barbels, scales, or the adipose fin \cite{Fishvista, AutoFish}. These challenges are further compounded by water turbidity, variable illumination, diverse body poses, and developmental stages. Accurate aquatic species recognition is therefore critical for applications including fisheries regulation \cite{garcia2022deepfish}, biodiversity monitoring \cite{saleh2020deepfish}, aquaculture management, invasive species detection, and climate-related studies. From underwater tracking systems for endangered species to intelligent aquaculture and illegal fishing surveillance, reliable species identification remains essential for modern aquatic monitoring systems \cite{10695087, 8478164, track_2}.

Given these challenges, relying solely on a single classification model is insufficient for fine-grained recognition \cite{Fishvista, saleh2020deepfish}. Instead, the problem naturally decomposes into three interrelated tasks grounded in biological reasoning: 1) \textit{semantic segmentation} of anatomical structures, 2) \textit{morphological trait prediction} of discriminative characteristics (\textit{e.g.}, fin shapes, stripe patterns, barbel arrangements), and 3) \textit{species classification} based on aggregated traits. These tasks exhibit a hierarchical dependency: fine-grained part segmentation provides the basis for inferring traits, while traits serve as high-level cues for species recognition. This bottom-up relationship can be formalized as
\begin{equation}\label{eq-bottom-up}
    P(y, \boldsymbol{t}, \boldsymbol{M} | \boldsymbol{X}) = \underbrace{P(y | \boldsymbol{t}, \boldsymbol{X})}_{\textit{Species}} \cdot \underbrace{P(\boldsymbol{t} | \boldsymbol{M}, \boldsymbol{X})}_{\textit{Traits}} \cdot \underbrace{P(\boldsymbol{M} | \boldsymbol{X})}_{\textit{Segmentation}}
\end{equation}
 where \(y\) is the species label, \(\boldsymbol{t}\) denotes the vector of morphological traits, \(\boldsymbol{M}\) is the dense segmentation mask, and \(\boldsymbol{X}\) is the input image. This formulation highlights how low-level morphological details progressively inform higher-level semantic understanding.

\subsection{Motivation}
\ph{From \eqref{eq-bottom-up}, effectively modeling this hierarchy remains challenging in practice. Most existing multi-task learning (MTL) frameworks \cite{FishNet, saleh2020deepfish, garcia2022deepfish} employ a shared backbone jointly optimized by multiple task objectives. However, segmentation, trait prediction, and species classification differ substantially in supervision density, loss scale, and optimization difficulty. Their gradients may therefore compete in the shared representation, causing high-level classification objectives to interfere with the fine-grained features required by lower-level morphological tasks. This \textit{negative transfer} is particularly detrimental to aquatic recognition, where species discrimination often depends on subtle anatomical structures. The key challenge is thus to preserve the hierarchical task dependencies while controlling both cross-task gradient interference and imbalance among heterogeneous task objectives.}

\ph{Despite recent progress, these two issues are not jointly addressed by existing approaches. Conventional MTL methods \cite{MTL_1, MTL_2} mainly rely on loss balancing or task-specific heads, without explicitly protecting lower-level representations from higher-level gradient interference. Gradient detachment has been effective in other learning paradigms \cite{SimSiam}, but has rarely been explored for enforcing hierarchical task dependencies in fine-grained recognition. Prototype-based segmentation can provide compact and discriminative representations \cite{Zhu_2025_CVPR}, but is typically studied in isolation from hierarchical MTL. Meanwhile, uncertainty weighting (UW) \cite{8578879} adaptively balances heterogeneous task losses according to their learned uncertainty, making it suitable for regulating how strongly segmentation, trait, and species objectives update the shared backbone. However, UW alone does not control where gradients propagate across task-specific representations and therefore cannot directly prevent hierarchical negative transfer. These limitations motivate a unified design that jointly regulates gradient propagation, task contribution, and morphological representation.}

\subsection{Main Contributions}
\begin{figure}[t]
    \centering
    \includegraphics[width=0.95\linewidth]{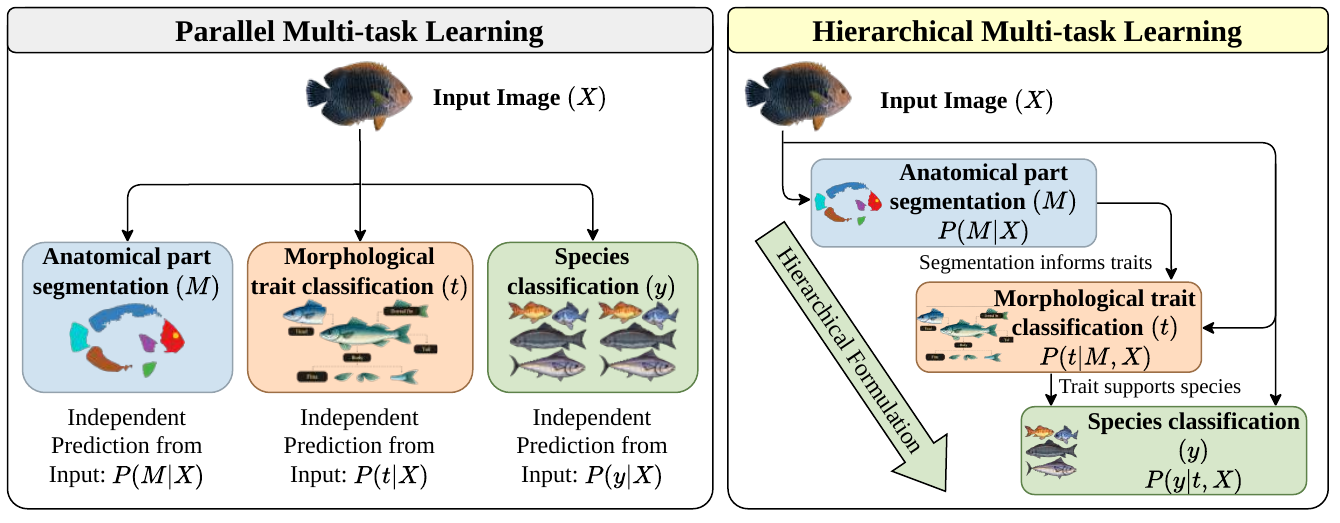}
    \caption{Comparison of parallel and hierarchical multi-task learning. Parallel MTL learns segmentation ($\boldsymbol{M}$), traits ($\boldsymbol{t}$), and species ($y$) independently from shared features, while $\FISHER$ models a hierarchical dependency where segmentation informs traits and traits guide classification.}
    \label{fig:parallel_vs_hierarchical_fisher}
\end{figure}

\ph{To address the gaps above, we propose $\FISHER$ (Fine-grained Integrated Segmentation and Hierarchical Learning for Recognition), a gradient-decoupled hierarchical framework for fine-grained aquatic species recognition that aligns multi-task optimization with the biological dependency among anatomical parts, morphological traits, and species. As illustrated in Fig.~\ref{fig:parallel_vs_hierarchical_fisher}, rather than learning these tasks in parallel, $\FISHER$ organizes them into a segmentation $\to$ trait $\to$ species hierarchy with asymmetric information and gradient flows. In the forward direction, \textit{segmentation} cues guide trait prediction, and the resulting trait representations support species classification. In the backward direction, gradients from higher-level objectives are detached at task boundaries, preventing them from directly modifying lower-level task-specific representations while allowing all tasks to jointly optimize the shared backbone. To support this hierarchy, the \textit{segmentation} head employs learnable prototypes with orthogonality regularization to obtain compact and separated anatomical representations, while homoscedastic uncertainty weighting adaptively balances the heterogeneous task objectives during joint optimization.}

\ph{Importantly, the novelty of $\FISHER$ lies not in stop-gradient, prototype learning, or uncertainty weighting individually, but in their integration within a biologically motivated \textit{structural optimization principle}. Motivated by the hierarchical dependency formalized through the probabilistic factorization, $\FISHER$ explicitly separates forward knowledge transfer from backward gradient propagation: higher-level tasks exploit lower-level morphological evidence without directly backpropagating through the corresponding task-specific representations. Accordingly, stop-gradient controls \textit{where} gradients propagate, prototype learning structures \textit{how} morphological evidence is represented and transferred, and uncertainty weighting regulates \textit{how strongly} the heterogeneous tasks contribute to joint optimization. Our structural and component ablations further demonstrate the importance of this coordinated design for mitigating negative transfer in hierarchical fine-grained recognition.}

The main contributions of this work are as follows:
\begin{itemize}
    \item \ph{We identify and \textit{directly measure} gradient conflict between dense morphological segmentation and global species classification as a fundamental source of negative transfer in fine-grained aquatic recognition, providing gradient-alignment evidence rather than accuracy alone.}
    \item \ph{We propose $\FISHER$, a gradient-decoupled hierarchical multi-task framework whose seg $\to$ trait $\to$ species dependency is derived from the posterior factorization. Specifically, we introduce boundary-level detachment to protect low-level representations from high-level interference while allowing the shared backbone to benefit from all supervision. Further, we demonstrate the necessity of this design, as uncertainty weighting alone degrades performance and generic gradient-surgery methods underperform on unseen species.}
    \item \ph{We introduce a \textit{prototype-based head} with orthogonality constraints, enabling precise, disentangled, and interpretable modeling of fine-grained anatomical structures.}
    \item \ph{We conduct extensive experiments on the large-scale Fish-Vista dataset and cross-domain validation on CUB-200-2011, demonstrating that $\FISHER$ achieves strong and consistent performance across semantic segmentation, trait prediction, and species classification tasks.}
\end{itemize}

The remainder of this paper is organized as follows. Section~\ref{sec:related} reviews related work. Section~\ref{sec:formulation} introduces the task formulation and architecture, and Section~\ref{sec:method} presents the proposed $\FISHER$ framework. Section~\ref{sec:experiments} describes the experimental setup and reports the results. Finally, Section~\ref{sec:conclusion} concludes the paper and discusses future directions. Key notations are summarized in Table~\ref{tab:notations}.

%------------------------ Table I -------------------------
\begin{table}[]
    \centering
    \caption{Notations and Symbols}
    \label{tab:notations}
    \renewcommand{\arraystretch}{1.2} 
    \resizebox{\columnwidth}{!}{

        \begin{tabular}{l p{7.5cm}} 
        \toprule
        Symbol & Description \\
        \midrule

        \multicolumn{2}{l}{\textit{Inputs and Targets}} \\
        \midrule
        $\boldsymbol{x} \in \mathbb{R}^{3 \times H \times W}$ & Input image \\
        $\boldsymbol{y}_s$ \& $ \hat{\boldsymbol{y}}_s\in \mathbb{R}^{K \times H \times W}$ & Ground-truth segmentation map and predicted logits \\
        $\boldsymbol{y}_t$ \& $ \hat{\boldsymbol{y}}_t\in \mathbb{R}^{T}$ & Ground-truth trait vector and predicted logits \\
        $\boldsymbol{y}_c$ \& $ \hat{\boldsymbol{y}}_c\in \mathbb{R}^{C}$ & Ground-truth species label and predicted logits\\
        $K, T$ \& $ C$ & Number of semantic classes, morphological traits, and species classes, resp. \\
        \midrule
        
        % --- Architecture Components ---
        \multicolumn{2}{l}{\textit{Architecture Components}} \\
        \midrule
        $\mathcal{F}$ \& $ \{\boldsymbol{f}_l\}$ & Feature extractor and hierarchical multi-scale feature map \\
        $\boldsymbol{f}$ & Fused high-resolution feature map via FPN \\
        $\boldsymbol{P} \in \mathbb{R}^{K \times D}$ & Learnable prototype matrix for segmentation \\
        $\boldsymbol{e}$ & Projected feature embedding for prototype matching \\
        $\boldsymbol{g}$ & Global image descriptor from the deepest backbone layer \\
        $\boldsymbol{p}$ & Aggregated segmentation evidence vector \\
        $\boldsymbol{b}$ & Base context vector (concatenation of $\boldsymbol{g}$ and detached $\boldsymbol{p}$) \\
        $\boldsymbol{b}_c$ & Extended context vector for species classification \\
        $\boldsymbol{h}_t$ \& $ \boldsymbol{h}_c$ & Hidden representations for trait and species heads \\
        $\psi_t$ \& $ \psi_c$ & MLP projectors for trait and species tasks \\
        \midrule
        
        % --- Loss Functions and Parameters ---
        \multicolumn{2}{l}{\textit{Loss Functions and Parameters}} \\
        \midrule
        $\mathcal{L}_s, \mathcal{L}_t$ \& $ \mathcal{L}_c$ & Task-specific losses for segmentation, traits, and species, resp. \\
        $\mathcal{L}_\mathtt{ortho}$ & Orthogonality regularization loss for prototypes \\
        $\eta_m$ (with $m \in \{s, t, c\}$) & Learnable parameter for uncertainty weighting  \\
        $\lambda_{ortho}$ & Hyperparameter balancing the orthogonality loss \\
        $\tau$ & Temperature scaling parameter for cosine similarity \\
        $\epsilon$ & Label smoothing factor \\
        \bottomrule
        \end{tabular}
    } \vspace{-1em}
\end{table}

\section{Related Work}
\label{sec:related}

\subsection{From Closed-Set Taxonomy to Open-World Analysis}
The field of aquatic visual analysis has evolved from closed-set classification to more complex open-world scenarios. Benchmarks such as WildFish++ \cite{zhuangwildfish} and AutoFish \cite{AutoFish}, which provide high-quality instance segmentation masks, have advanced fine-grained categorization. Similarly, FishNet \cite{FishNet} extended this paradigm by incorporating functional trait prediction across over $17,000$ species.
Recent progress has further improved performance by transitioning from Convolutional Neural Networks (CNNs) to Vision Transformers (ViTs). For example, Veiga and Rodrigues \cite{10636215} demonstrated that Swin Transformers with fine-grained modules achieve state-of-the-art (SOTA) results, while Tejaswini \textit{et al.} \cite{10695087} showed that ViT-B16 outperforms CNNs for estuarine species classification.

Despite these advances, most approaches, including lightweight models such as YOLOv8n-DFG \cite{ren_2025} and foundation models like BioCLIP \cite{stevens2024bioclip}, rely heavily on global feature aggregation. While effective for frequent classes, this strategy struggles with \textit{ultra-rare species}, where global representations are difficult to learn from limited data. In contrast, we argue that explicit morphological segmentation provides a more robust and invariant prior for recognizing rare species.

\subsection{MTL: Shared Representations and Gradient Conflicts}
MTL aims to improve generalization by exploiting shared representations across related tasks. In aquatic vision, existing frameworks often integrate multiple objectives; for example, YOLO-FD \cite{lizhao_2024} combines detection and segmentation for fish disease analysis, while DeepSeaVision \cite{11193787} enhances YOLOv9 with image enhancement techniques (\textit{e.g.}, CLAHE, UCM) for underwater environments. Other works, such as \cite{10895375}, explore hybrid models to preserve low-level features that are often lost in deep architectures.
However, shared backbones are inherently limited by negative transfer, where conflicting task objectives degrade performance \cite{standley2020tasks}. This issue is related to challenges in the model-protected MTL \cite{liang2022}, where shared parameters can introduce unintended cross-task interference. Moreover, recent studies on long-tailed learning \cite{xiong_man_lv_xu_zeng_shi_lai_yang_2026} suggest that effective learning often requires decoupling representation learning from classification.

\ph{Building on this insight, we address the imbalance between dominant global features and subtle local morphological traits. Unlike prior methods that rely on complex gradient manipulation (\textit{e.g.}, PCGrad \cite{yu2020gradient}), $\FISHER$ introduces a structural solution through hierarchical gradient decoupling, preventing high-level semantic gradients from interfering with low-level anatomical feature learning. It further regularizes segmentation prototypes through an orthogonality constraint. Unlike methods such as D2MoRA~\cite{zuo2026d2mora}, which enforce orthogonality in the expert-parameter space, our approach applies it in the prototype (part) space to separate fine-grained anatomical representations.}

\subsection{Hierarchical and Interpretable Feature Learning}
Hierarchical learning has emerged as an effective strategy for handling data scarcity and label ambiguity. Prior work demonstrated that linking coarse body structures with fine-grained features improves fish recognition \cite{yin_2023}, while hierarchical attention mechanisms help reduce misleading feature focus in fine-grained classification \cite{10854460}.

$\FISHER$ extends these ideas by explicitly modeling a biologically grounded dependency chain: \textit{segmentation} (anatomy) informs \textit{traits} (attributes), which in turn guide \textit{species} recognition. Unlike Concept Bottleneck Models (CBMs) \cite{koh2020concept} or weakly supervised attention approaches, our framework enforces structured dependencies through supervised, prototype-based representations. This design improves interpretability and leverages biological priors to learn meaningful feature hierarchies, leading to strong generalization on ultra-rare species, as demonstrated on Fish-Vista \cite{Fishvista}.

%%%%%%%%%%%%%%%%%%%%%%%%%%%
\section{Task Formulation and Architecture}
\label{sec:formulation}
\begin{figure*}[t]
    \centering
   \includegraphics[width=1\linewidth]{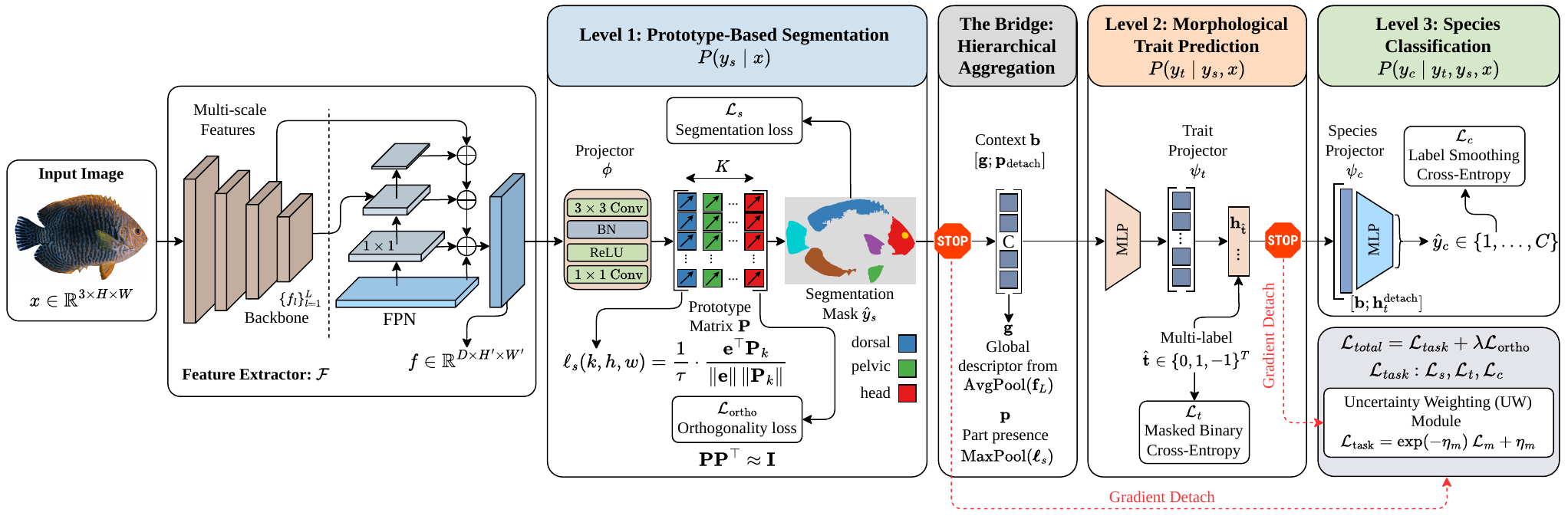} \vspace{-1.0em}
    \caption{Overview of the $\FISHER$ architecture. Features are extracted by a shared backbone and refined hierarchically from segmentation to trait prediction and species classification. {\color{red}Red} dashed lines denote gradient decoupling (stop-gradient) used to mitigate negative transfer.}
    \label{fig:overall_framework}
\end{figure*}

In this section, we present the core components of a detached hierarchical multi-task framework with learnable prototypes. We first formalize the hierarchical task formulation using a probabilistic hierarchy, then describe the $\FISHER$ architecture.

\subsection{Hierarchical Task Formulation}
\label{sec:problem}
Fine-grained fish recognition is challenging due to subtle inter-class differences and intra-class variations in morphological attributes. Given an input image \(\boldsymbol{x} \in \mathbb{R}^{3 \times H \times W}\) with \(H\) and \(W\) being the height and width of the image, our goal is to jointly learn three related tasks:
\begin{itemize}
\item \textit{Semantic segmentation}: A per-pixel label map \(\boldsymbol{y}_s \in \{0, 1, \dots, K-1\}^{H \times W}\), where \(K\) is the number of semantic classes (\textit{e.g.}, {fins}, {head}, and {body} regions). The label 0 is reserved for \textit{background} or \textit{ignored} regions.
\item \textit{Morphological trait prediction}: A multi-label binary vector \(\boldsymbol{y}_t \in \{0, 1, -1\}^T\), where \(T\) is the number of traits (\textit{e.g.}, {shape of caudal fin}), and \(-1\) indicates {missing} or {unavailable} annotations in partial labels.
\item \textit{Species classification}: A categorical label \(\boldsymbol{y}_c \in \{1, \dots, C\}\), where \(C\) is the number of species.
\end{itemize}
\ph{To avoid ambiguity, we use two terms consistently throughout the paper. A \textit{part} denotes an anatomical region produced by the segmentation task (\textit{e.g.}, \textit{dorsal fin}, \textit{head}, \textit{body}); the learnable prototypes $\boldsymbol{P}$ correspond to \textit{parts}. A \textit{trait} denotes a morphological attribute predicted by the trait head (\textit{e.g.}, \textit{shape of the caudal fin}, \textit{presence of barbels}). Thus, parts are spatial (pixel-level) entities, whereas traits are image-level attributes inferred from them.}

These tasks exhibit a natural hierarchical dependency rooted in biological semantics: precise \textit{segmentation} of anatomical parts provides foundational evidence for inferring \textit{traits}; then, the inferred \textit{traits} serve as discriminative cues for \textit{species identification}. This hierarchical dependency can be formally expressed through the following joint posterior factorization: 
{\small\begin{equation} \label{eq:1}
P(\boldsymbol{y}_s, \boldsymbol{y}_t, \boldsymbol{y}_c |  \boldsymbol{x}) = P(\boldsymbol{y}_s |  \boldsymbol{x}) \cdot P(\boldsymbol{y}_t |  \boldsymbol{y}_s, \boldsymbol{x}) \cdot P(\boldsymbol{y}_c |  \boldsymbol{y}_t, \boldsymbol{y}_s, \boldsymbol{x})
\end{equation}}where \(P(\boldsymbol{y}_s | \boldsymbol{x})\) captures dense  pixel-level morphological details, \(P(\boldsymbol{y}_t |  \boldsymbol{y}_s, \boldsymbol{x})\) aggregates these into trait-level representations, and \(P(\boldsymbol{y}_c |  \boldsymbol{y}_t, \boldsymbol{y}_s, \boldsymbol{x})\) leverages the aggregated information to clarify global categorization.

\ph{Although species identity can, in principle, provide useful information for trait inference, allowing the species objective to directly shape lower-level task-specific representations may bias them toward species-specific correlations, thereby reducing their transferability to unseen categories. Since our goal is to generalize to ultra-rare and unseen species, we instead adopt the bottom-up hierarchy as an \textit{optimization prior}, allowing morphological evidence (\textit{e.g.} fin shapes) to propagate toward species recognition while preventing higher-level gradients from directly modifying lower-level task-specific representations. This design encourages transferable morphological representations while retaining joint supervision of the shared backbone. The effects of reversing or fully connecting this optimization flow are further evaluated in the ablation study.}

\subsection{Overall Architecture}\label{sec:formulation-B}
As illustrated in Fig. \ref{fig:overall_framework}, the proposed $\FISHER$ framework consists of a shared feature extractor \(\mathcal{F}\) that processes the input
 \(\boldsymbol{x}\) into multi-scale feature maps \(\{\boldsymbol{f}_l\}_{l=1}^L\), where \(\boldsymbol{f}_l \in \mathbb{R}^{C_l \times H_l \times W_l}\) represents the feature map at level \(l\), with \(C_l\) channels and spatial dimensions \(H_l \times W_l\) (typically decreasing with \(l\)).
These multi-scale features are fused using a lightweight Feature Pyramid Network (FPN) to generate a high-resolution, semantically rich feature map \(\boldsymbol{f} \in \mathbb{R}^{D \times H' \times W'}\), where \(D\) is the unified channel dimension, and \(H'\) and \(W'\) are downsampled spatial dimensions. The FPN employs lateral convolutions ($1\times1$) to align channels and top-down upsampling with additions, followed by smoothing convolutions ($3\times3$).

From \(\boldsymbol{f}\), the model branches hierarchically into three specialized heads:
\begin{mdframed}
\renewcommand{\labelenumi}{(\arabic{enumi})}
\begin{enumerate}
\item[\circled{\cnt}] First, the segmentation head processes \(\boldsymbol{f}\) to produce \textit{segmentation} logits \(\hat{\boldsymbol{y}}_s\).

\item[\circled{\cnt}] Next, the \textit{trait} head aggregates \textit{segmentation} evidence detached from gradient propagation, together with global image features to infer morphological traits.

\item[\circled{\cnt}] Finally, the \textit{species} head utilizes detached \textit{trait} embeddings along with shared contextual features to perform species classification.
\end{enumerate}
\end{mdframed}
To enforce a unidirectional information flow, gradient detachment is applied at the interfaces between successive heads, preventing gradients from propagating from higher-level tasks to lower-level representations and thereby preserving fine-grained morphological features. \ph{Concretely, this detachment isolates the downstream classification gradients from the intermediate segmentation logits ($\boldsymbol{l}_s$), while the shared backbone ($\boldsymbol{f}_L$) is still co-optimized by all tasks via the non-detached global descriptor $\boldsymbol{g}$.}

\section{Methodology}
\label{sec:method}
We now detail the three specialized heads, including the prototype-based \textit{segmentation} head, gradient-detached propagation for \textit{morphological trait prediction} and \textit{species classification}, as well as the loss functions and training procedure.

\subsection{Prototype-Based Semantic Segmentation}
We adopt a learnable prototype-based head to replace conventional convolutional decoders for modeling morphologically accurate part representations. Instead of dense decoding, semantic classes are represented as compact vectors in feature space, enabling clearer separation and more robust segmentation. \ph{We choose prototypes over a dense decoder for four task-specific reasons. First, similarity-based assignment to class prototypes provides part-specific representations, particularly for small and visually similar structures (\textit{e.g.}, \textit{barbels} and \textit{adipose fin}). Second, each prototype serves as a learnable morphological reference, whose per-class evidence (Section~\ref{sec:formulation-B}) provides a compact signal to the trait head. Third, using $K$ prototype vectors reduces parameter overhead compared with a heavy dense decoder. Fourth, the formulation supports orthogonality regularization that promotes separation among similar anatomical parts, as illustrated in Figs.~\ref{fig:tsne} and~\ref{fig:prototype_corr}.} Let \(\boldsymbol{P} \in \mathbb{R}^{K \times D}\) denote the learnable prototype matrix, where each row \(\boldsymbol{P}_k \in \mathbb{R}^D\) corresponds to class \(k \in \{0, \dots, K-1\}\). The prototypes are initialized orthogonally to promote initial diversity.

The head first projects the FPN feature \(\boldsymbol{f}\) into a prototype-aligned embedding space:
\begin{equation}
\boldsymbol{e} = \phi(\boldsymbol{f}) \in \mathbb{R}^{D \times H' \times W'}
\end{equation}
where \(\phi\) consists of a stacked sequence $\mathtt{Conv2D} $($3\times3$), $\mathtt{BatchNorm2D}$, $\mathtt{ReLU}$, followed by $\mathtt{Conv2D}$ ($1\times1$). For each pixel position \((h, w)\), the \textit{segmentation} logits are computed using normalized cosine similarity scaled by a temperature parameter \(\tau > 0\), as follows:
\begin{equation}
\boldsymbol{l}_s(k, h, w) = \frac{1}{\tau} \cdot \frac{\boldsymbol{e}(\cdot, h, w)^\top \boldsymbol{P}_k}{\|\boldsymbol{e}(\cdot, h, w)\|_2 \cdot \|\boldsymbol{P}_k\|_2}
\end{equation}
yielding \(\boldsymbol{l}_s \in \mathbb{R}^{K \times H' \times W'}\). The final segmentation logits are obtained via bilinear upsampling
\begin{equation}
\hat{\boldsymbol{y}}_s = \mathtt{Upsample}(\boldsymbol{l}_s) \in \mathbb{R}^{K \times H \times W}.
\end{equation}

To ensure distinct and discriminative prototypes, we impose an orthogonality regularization as
\begin{equation}
\mathcal{L}_{\mathtt{ortho}} = \frac{1}{K^2} \big\| \boldsymbol{P}_\mathtt{norm} \boldsymbol{P}_\mathtt{norm}^\top - \boldsymbol{I}_K \big\|_F^2
\label{eq:ortho_loss}
\end{equation}
 where \(\boldsymbol{I}_K\) is the identity matrix, \(\boldsymbol{P}_\mathtt{norm}\) is the row-wise \(\ell_2\)-normalized prototype matrix, and \(\|\cdot\|_F\) represents the Frobenius norm. 
 By reducing similarity among prototypes, this loss enforces pairwise orthogonality and promotes a well-separated feature space for capturing subtle morphological variations. For datasets with a background class, orthogonality can be applied only to the 
$K-1$ foreground prototypes to avoid constraining non-semantic regions.
This prototype-based formulation is particularly advantageous for \textit{fine-grained segmentation}: it naturally handles class imbalance through similarity-based assignment, provides interpretable representations (\textit{e.g.}, prototypes as class centroids), and reduces parameter overhead compared to dense decoders.

\subsection{Hierarchical Aggregation}
To propagate hierarchical information while respecting the conditional dependencies, we aggregate context from lower levels for higher tasks. First, a global image descriptor is derived from the deepest backbone feature \(\boldsymbol{f}_L \in \mathbb{R}^{C_L \times H_L \times W_L}\)
\begin{equation}
\boldsymbol{g} = \mathtt{AdaptiveAvgPool2D}(\boldsymbol{f}_L) \in \mathbb{R}^{C_L}
\end{equation}
capturing holistic scene information. Segmentation evidence is summarized as the maximum activation across spatial dimensions for each prototype channel:
\begin{equation}
\boldsymbol{p} = \mathtt{AdaptiveMaxPool2D}(\boldsymbol{l}_s) \in \mathbb{R}^K.
\end{equation}
This vector represents the strength of presence for each semantic class in the image. 
The base context vector is then formed by concatenation
\begin{equation}
\boldsymbol{b} = [\boldsymbol{g}; \boldsymbol{p}^\mathtt{detach}] \in \mathbb{R}^{C_L + K}
\end{equation}
where \(\boldsymbol{p}^\mathtt{detach}\) denotes the detached tensor ({gradient flow stopped}), ensuring that downstream task {trait} or {classification} optimization does not influence {segmentation} parameters. 

\subsection{Detached Downstream Tasks}
For morphological trait prediction, the context vector \(\boldsymbol{b}\) is fed into a \textit{trait} projection module \(\psi_t: \mathbb{R}^{C_L + K} \to \mathbb{R}^{d_t}\), implemented as a multi-layer perceptron (MLP) with normalization and non-linear activation, as follows:
\begin{equation}
\boldsymbol{h}_t = \psi_t(\boldsymbol{b}) \in \mathbb{R}^{d_t}.
\end{equation}
The trait logits are then computed as
\begin{equation}
\hat{\boldsymbol{y}}_t = \boldsymbol{W}_t \boldsymbol{h}_t + \boldsymbol{b}_t \in \mathbb{R}^T
\label{traitlogits}
\end{equation}
where \(\boldsymbol{W}_t \in \mathbb{R}^{T \times d_t}\) is the weight matrix and \(\boldsymbol{b}_t \in \mathbb{R}^T\) is the bias vector of the final linear projection. To preserve the morphological fidelity captured by \(P(\boldsymbol{y}_s |  \boldsymbol{x})\), gradient detachment is applied such that gradients from the trait loss do not propagate back to the segmentation branch, while segmentation-derived features still contribute to trait prediction during forward inference.

Building upon this representation, the species classification head further incorporates trait information within the hierarchical structure
\begin{equation}
\boldsymbol{b}_c = [\boldsymbol{b}; \boldsymbol{h}_t^\mathtt{detach}] \in \mathbb{R}^{C_L + K + d_t}
\end{equation}
\noindent where \(\boldsymbol{h}_t^\mathtt{detach}\) is detached to prevent \textit{species} gradients from flowing back to the trait head. This extended context is processed by a MLP species projector \(\psi_c: \mathbb{R}^{C_L + K + d_t} \to \mathbb{R}^{d_c}\), as follows
\begin{equation}
\boldsymbol{h}_c = \psi_c(\boldsymbol{b}_c) \in \mathbb{R}^{d_c}.
\end{equation}
The final species logits are computed as
\begin{equation}
\hat{\boldsymbol{y}}_c = \boldsymbol{W}_c \boldsymbol{h}_c + \boldsymbol{\beta}_c \in \mathbb{R}^C
\label{classlogits}
\end{equation}
where \(\boldsymbol{W}_c \in \mathbb{R}^{C \times d_c}\) is the weight matrix and \(\boldsymbol{\beta}_c \in \mathbb{R}^C\) is the bias vector of the final linear projection.

This hierarchical detachment strategy ensures that optimization of the species task does not bias intermediate trait representations, while the shared context 
\(\boldsymbol{b}\) enables the backbone to benefit from supervision across all tasks.

\subsection{Loss Functions and Task Balancing}
The model is optimized using task-specific loss functions, which are dynamically balanced to mitigate the dominance of easier tasks and ensure stable multi-task learning.

\noindent \textbf{Segmentation}: We employ a combination of weighted cross-entropy and Dice loss to address class imbalance and improve overlap accuracy. First, the predicted probability \(p_{i,k}\) for pixel \(i\) and class \(k\) is obtained via the softmax function:
\begin{equation}
p_{i,k} = \frac{\exp\!\left(\hat{y}_{s,i,k}\right)}{\sum_{j=0}^{K-1}\exp\!\left(\hat{y}_{s,i,j}\right)}
\label{probseg}
\end{equation}
where \(\hat{y}_{s,i,k}\) is the segmentation logit for pixel \(i\) and class \(k\). The weighted cross-entropy is then defined as:
\begin{equation}
\mathcal{L}_s^{\mathrm{CE}} = -\frac{1}{\sum_{i\in\Omega} w_{y_{s,i}}} \sum_{i\in\Omega} w_{y_{s,i}} \log (p_{i,y_{s,i}})
\end{equation}
where \(\Omega\) is the set of valid pixels, \(y_{s,i} \in \{0, \dots, K-1\}\) is the ground-truth class index for pixel \(i\), \(w_k\) is the class weight computed as inverse normalized frequencies, and \(p_{i,y_{s,i}}\) is the predicted probability corresponding to the ground-truth class \(y_{s,i}\). To further encourage region-level consistency, we incorporate the Dice loss, which promotes intersection-over-union
\begin{equation}
\mathcal{L}_s^\text{Dice} = 1 - \frac{1}{K} \sum_{k=0}^{K-1} \frac{2 \sum_{i \in \Omega} p_{i,k} \cdot g_{i,k}}{\sum_{i \in \Omega} p_{i,k} + \sum_{i \in \Omega} g_{i,k}}
\end{equation}
where  \(g_{i,k} \in \{0, 1\}\) is the one-hot encoded ground truth for pixel \(i\) and class \(k\). The total segmentation loss is computed as follows:
\begin{equation}
\mathcal{L}_s = \mathcal{L}_s^\text{CE} + \mathcal{L}_s^\text{Dice}.
\end{equation}

\noindent \textbf{Morphological trait prediction}: We use a masked binary cross-entropy loss to accommodate partial labels, formulated as follows:
\begin{align}
\mathcal{L}_t
&=
-\frac{1}{|\mathcal{M}|}
\sum_{j \in \mathcal{M}}
\Big[
y_{t,j} \log \sigma(\hat{y}_{t,j}) \nonumber\\
&\qquad\qquad\qquad
+ (1 - y_{t,j}) \log \big(1 - \sigma(\hat{y}_{t,j})\big)
\Big]
\end{align}
where \(\mathcal{M} = \{j \mid y_{t,j} \neq -1\}\) is the set of valid trait indices. Here, \(y_{t,j} \in \{0, 1\}\) is the ground-truth binary label for the \(j\)-th trait, and \(\hat{y}_{t,j}\) is its corresponding predicted logit (\textit{i.e.}, the \(j\)-th element of the vector \(\hat{\boldsymbol{y}}_t\) defined in \eqref{traitlogits}). The function \(\sigma(z) = 1 / (1 + e^{-z})\) denotes the sigmoid activation.

\noindent \textbf{Species classification}: We employ cross-entropy loss with label smoothing given by
\begin{equation}
\mathcal{L}_c = -\sum_{k=1}^C \tilde{y}_{c,k} \log (p_{c,k})
\end{equation}
where \(p_{c,k}\) is the predicted probability for class \(k\) obtained by applying the softmax function to the species logit vector \(\hat{\boldsymbol{y}}_c \in \mathbb{R}^C\). The smoothed ground-truth target \(\tilde{y}_{c,k}\) is defined as
\begin{equation}
\tilde{y}_{c,k} = 
\begin{cases} 
1 - \epsilon, & \text{if } k = y_c \\ 
\frac{\epsilon}{C-1}, & \text{otherwise}
\end{cases}
\end{equation}
where \(y_c \in \{1, \dots, C\}\) is the ground-truth species class index, and \(\epsilon\) is the label smoothing factor. This formulation encourages the model to produce softer probability distributions, reducing overconfidence and improving generalization, which is particularly beneficial in fine-grained classification tasks with potential label noise or ambiguity.

To balance the contributions of different tasks, we adopt uncertainty weighting \cite{8578879}, which addresses imbalance in gradient magnitudes:
\begin{equation}
\label{uw_loss}
\mathcal{L}_\mathtt{task} = \sum_{m \in \{s,t,c\}} \big( \exp(-\eta_m) \mathcal{L}_m + \eta_m \big)
\end{equation}
where \(m\) indexes the three tasks (\textit{i.e.}, segmentation \(s\), trait prediction \(t\), and species classification \(c\)), and \(\eta_m = \log (\sigma_m^2)\) are task-specific learnable parameters. Intuitively, this formulation adaptively down-weights noisy or difficult tasks while preventing degenerate solutions through the additive \(\eta_m\) term. \ph{This mechanism is particularly suitable for our heterogeneous multi-task setting, where segmentation, trait prediction, and long-tailed species classification exhibit different loss scales and optimization difficulties. It complements gradient detachment by regulating \emph{how strongly} each task updates the shared backbone, while detachment determines \emph{where} gradients are allowed to propagate.}

\textbf{The training objective}: The overall training objective of the proposed framework consists of multiple task-specific loss terms in \eqref{uw_loss} together with an orthogonality regularization term in \eqref{eq:ortho_loss}, which is formulated as
\begin{equation}
\mathcal{L}_\mathtt{total} = \mathcal{L}_\mathtt{task} + \lambda_{ortho} \mathcal{L}_\mathtt{ortho}
\label{totalloss}
\end{equation}
where \(\lambda_{ortho}\) controls the strength of the regularization. Together, these losses enable robust optimization by improving segmentation boundary precision, supporting partial supervision, and maintaining classification accuracy.

\section{Experiments}
\label{sec:experiments}
In this section, we evaluate the proposed $\FISHER$ framework. We first describe the experimental setup, followed by comparisons with SOTA methods on the Fish-Vista benchmark. We then present results across three tasks, including \textit{semantic segmentation}, \textit{trait identification}, and \textit{species classification}, highlighting the effectiveness of the decoupled learning strategy under long-tailed distributions.

\subsection{Experimental Setup}

\noindent\textbf{Dataset}: We evaluate our method primarily on the Fish-Vista dataset~\cite{Fishvista}, the largest and most diverse benchmark for fine-grained aquatic trait analysis. It comprises $69,269$ images covering $4,316$ fish species, annotated with hierarchical labels (\textit{species} taxonomy, \textit{trait} presence, and pixel-level \textit{segmentation} masks). We adhere to the official evaluation protocol using three test splits:
\begin{itemize}
    \item (1) \textit{In-Species} (seen species),
    \item (2) \textit{Leave-Out-Species} (unseen species, testing Out-Of-Distribution generalization), and
    \item (3) \textit{Manual-Annotation} (high-quality labels verified by experts).
\end{itemize}

\ph{In the In-Species and Leave-Out-Species splits, trait labels are \textit{propagated from species-level taxonomic databases}, whereas the Manual-Annotation split uses image-level labels obtained through \textit{direct visual inspection by human experts}~\cite{Fishvista}.}

\ph{We further verify the proposed method on CUB-200-2011~\cite{wah2011caltech} to assess cross-domain generalization, using its $312$ binary attributes as trait supervision and species labels for classification. Since CUB lacks dense anatomical masks, we construct coarse part-region masks by drawing a fixed-radius disk around each of the $15$ annotated part keypoints and use them only as auxiliary segmentation supervision. As these pseudo masks are geometric approximations rather than expert annotations, segmentation mIoU is therefore not reported, and the evaluation focuses on species classification and attribute prediction. To examine severe class imbalance, CUB is additionally resampled into an ultra-rare long-tailed setting (CUB-LT) with an imbalance factor of $500{:}1$ and $96$ tail classes. Detailed results are reported in Sec.~\ref{sec:cub}.}

\vspace{5pt}\noindent\textbf{Evaluation metrics}:
To evaluate the proposed framework across hierarchical tasks, we adopt task-specific metrics tailored to each prediction objective. For semantic segmentation, we use mean Intersection-over-Union (mIoU), a standard metric that accounts for pixel-wise accuracy under class imbalance. For each class \(k \in \{1, \dots, K-1\}\), the IoU is defined as
\begin{equation}
\mathtt{IoU}_k = \frac{\mathtt{TP}_k}{\mathtt{TP}_k + \mathtt{FP}_k + \mathtt{FN}_k}
\end{equation}
where \(\mathtt{TP}_k\), \(\mathtt{FP}_k\), and \(\mathtt{FN}_k\) denote true positives, false positives, and false negatives for class \(k\), respectively. The mIoU is then computed as the average across classes:
\begin{equation}
\mathtt{mIoU} = \frac{1}{K-1} \sum_{k=1}^{K-1} \mathtt{IoU}_k.
\end{equation}
This metric evaluates the model’s ability to accurately delineate fine-grained anatomical regions, which are critical for downstream tasks.

For morphological trait prediction, given its multi-label binary nature, we adopt the macro-averaged F1-score to balance precision and recall across traits. For each trait \(j \in \{1, \dots, T\}\), precision and recall are determined as follows
\begin{align}
\mathtt{Precision}_j 
&= \frac{\mathtt{TP}_j}{\mathtt{TP}_j + \mathtt{FP}_j} \\
\mathtt{Recall}_j    
&= \frac{\mathtt{TP}_j}{\mathtt{TP}_j + \mathtt{FN}_j}
\end{align}
and the corresponding F1-score is given by:
\begin{equation}
\mathtt{F1}_j = \frac{2 \cdot \mathtt{Precision}_j \cdot \mathtt{Recall}_j}{\mathtt{Precision}_j + \mathtt{Recall}_j}.
\end{equation}

%------------------------ Table II -------------------------
\begin{table}[]
\centering
\textcolor{black}{
\caption{Frequency-stratified class distribution of Fish-Vista~\cite{Fishvista},
computed on the $47{,}595$-image \textit{train} split: \textit{Majority} ($\ge500$), \textit{Neutral}
($100$--$499$), \textit{Minority} ($10$--$99$), \textit{Ultra-Rare} ($<10$) images}
\label{tab:fishvista_stats}
\renewcommand{\arraystretch}{1.15}
\resizebox{\columnwidth}{!}{%
\begin{tabular}{l r r r r}
\toprule
Group & \#Classes & \% Classes & \#Train Images & \% Images \\
\midrule
Majority ($\ge500$)     & 20    & 0.6\%  & 25{,}246 & 53.0\% \\
Neutral ($100$--$499$)  & 35    & 1.0\%  & 7{,}278  & 15.3\% \\
Minority ($10$--$99$)   & 331   & 9.1\%  & 7{,}458  & 15.7\% \\
Ultra-Rare ($<10$)      & 3{,}233 & 89.3\% & 7{,}613  & 16.0\% \\
\midrule
Total (trained species) & 3{,}619 & 100\% & 47{,}595 & 100\% \\
\bottomrule
\end{tabular}%
}
}
\vspace{-1em}
\end{table}

%------------------------ Table III -------------------------
\begin{table}[]
\centering
\caption{Implementation Details and Hyperparameter Settings}
\label{tab:implementation_details}
\resizebox{\columnwidth}{!}{
% Thay đổi quan trọng nhất ở đây: {lc} thay vì {l|c}
\begin{tabular}{lc} 
\toprule
Configuration & Value/Setting \\
\midrule

\multicolumn{2}{l}{\textit{Optimization Setup}} \\
\midrule
Optimizer & AdamW \\
Learning rate & $1 \times 10^{-4}$ \\
Weight decay & $0.01$ \\
Batch size & 32 \\
Training epochs & 50 \\ 
LR scheduler & Cosine annealing \\
Warmup epochs & 5 \\
\midrule

\multicolumn{2}{l}{\textit{Model Architecture}} \\
\midrule
Backbone & Swin-Base (ImageNet-22k) \\
Input size & $224 \times 224$ \\
Prototype temperature ($\tau$) & 0.1 \\
Number of prototypes & 10 (Background + 9 Traits) \\
\midrule

\multicolumn{2}{l}{\textit{Loss Coefficients}} \\
\midrule
Orthogonality Weight ($\lambda_{ortho}$) & 0.5 \\
Label Smoothing ($\epsilon$) & 0.1 \\ 
\bottomrule
\end{tabular}
} \vspace{-1.0em}
\end{table}

In addition, we employ mean Average Precision (mAP) to evaluate ranking performance across traits. The Average Precision (AP) for each trait 
$j$ is defined as:
\begin{equation}
\mathtt{AP}_j = \sum_{k=1}^{N} (R_k - R_{k-1}) \cdot P_k
\end{equation}
\noindent where \(P_k\) and \(R_k\) denote precision and recall at the \(k\)-th threshold when ranking samples by the predicted score \(\hat{\boldsymbol{t}}_j\), and \(N\) is the number of samples. The mAP is then averaged over all traits
\begin{equation}
\mathtt{mAP} = \frac{1}{T} \sum_{j=1}^T \mathtt{AP}_j.
\end{equation}
This metric complements the macro-F1 score by evaluating the model’s ability to rank positive traits effectively.

For species classification, we use top-1 accuracy, which measures the proportion of correctly classified samples
\begin{equation}
\mathtt{Accuracy} = \frac{1}{N} \sum_{i=1}^N \mathbb{I}(\argmax_k \hat{y}_{c,i,k} = y_{c,i})
\end{equation}
\noindent where \(N\) is the number of samples, and \(\mathbb{I}(\cdot)\) is the indicator function. This straightforward metric evaluates the global categorization performance, benefiting from the hierarchical cues provided by lower-level tasks. Given the extreme class imbalance in Fish-Vista \cite{Fishvista}, we stratify performance into four frequency-based groups based on numbers of training samples: \textit{Majority} ($\ge 500$ images), \textit{Neutral} ($100-499$), \textit{Minority} ($10-99$), and \textit{Ultra-Rare} ($<10$). The \textit{Ultra-Rare} metric is of particular interest as it reflects the model's ability to learn from few-shot examples. \ph{Table~\ref{tab:fishvista_stats} details this distribution. The $20$ \textit{Majority} species account for over half of the training images, whereas the $3{,}233$ \textit{Ultra-Rare} species together contribute only $16$\%. A further $697$ species have no training images and are evaluated only under the Leave-Out-Species protocol.}

%------------------------ Table IV -------------------------
\begin{table}[]
\centering
\small
\begin{threeparttable}
\caption{Comparison of Classification Performance (in \%). Results Are Color-coded as \colorbox{blue!35}{Best}, \colorbox{blue!15}{Second best}, \colorbox{red!35}{Worst}, and \colorbox{red!15}{Second worst}}
\label{tab:classification}

\setlength{\tabcolsep}{2.5pt}
\renewcommand{\arraystretch}{1.1}

\begin{tabular}{lccccc}
    \toprule
    \multirow{2.1}{*}{{Model}} & \multirow{2.1}{*}{{F1}} & Major & Neutral & Minor & Ultra-R \\
    &  & Acc. & Acc. & Acc. & Acc. \\
    \midrule
    % --- CNNs ---
    VGG-19 \cite{simonyan2014very} & 49.7 & 93.5 & 83.0 & 74.2 & 45.9 \\
    ResNeXt-50 \cite{xie2017aggregated} & 44.4 & 91.4 & 78.3 & 69.8 & 39.1 \\
    RegNetY-4G \cite{radosavovic2020designing} & 43.7 & 89.8 & 77.4 & 68.5 & 38.5 \\
    \midrule
    % --- Transformers ---
    ViT-B16 \cite{ViT} & 48.3 & 88.7 & 82.3 & 73.3 & 43.4 \\
    Swin-B-22k \cite{swin} & 55.1 & 92.6 & 86.2 & 79.6 & 50.4 \\
    MaxViT-T \cite{maxvit} & 57.8 & 94.4 & \colorbox{blue!15}{86.7} & 81.4 & 53.9 \\
    \midrule
    % --- Foundation Models (Linear Probe) ---
    BioCLIP-LP \cite{stevens2024bioclip} & 38.2 & \colorbox{red!15}{75.5} & \colorbox{red!15}{65.2} & 61.3 & 31.1 \\
    CLIP-LP \cite{radford2021learning} & \colorbox{red!15}{25.4} & \colorbox{red!35}{55.9} & \colorbox{red!35}{49.8} & \colorbox{red!15}{46.7} & \colorbox{red!15}{20.9} \\
    DINOv2-LP \cite{oquab2023dinov2} & 53.1 & 89.9 & 78.0 & 76.0 & 47.0 \\
    \midrule
    % --- FGVC Methods ---
    INTR \cite{paul2023simple} & \colorbox{red!35}{6.1} & 92.2 & 73.2 & \colorbox{red!35}{22.6} & \colorbox{red!35}{0.6} \\
    TransFG \cite{he2022transfg} & 50.3 & 94.5 & 86.6 & 75.5 & 45.3 \\
    \midrule
    % --- Ours ---
    MTLSwinB & \colorbox{blue!15}{57.9} & \colorbox{blue!15}{95.2} & \colorbox{blue!35}{89.0} & \colorbox{blue!15}{85.9} & \colorbox{blue!15}{60.5} \\
    % AquaDeHi & 57.5 & 94.9 & \colorbox{blue!15}{89.0} & 84.1 & \colorbox{blue!15}{61.4} \\
    $\FISHER$ & \colorbox{blue!35}{60.8} & \colorbox{blue!35}{95.5} & \colorbox{blue!35}{89.0} & \colorbox{blue!35}{86.2} & \colorbox{blue!35}{63.8} \\
    \bottomrule
\end{tabular}

\end{threeparttable}
\vspace{-1.0em}
\end{table}

%------------------------ Table V -------------------------

\begin{table}[!ht]
\centering
\textcolor{black}{
\setlength{\fboxsep}{2pt}
\renewcommand{\arraystretch}{1.2}
\caption{Comparison with Specialized Long-tailed Methods on Fish-Vista (highlighted as \colorbox{blue!35}{Best}, \colorbox{blue!15}{Second best}, \colorbox{red!35}{Worst}, \colorbox{red!15}{Second worst}). All Long-tailed Rows Share $\FISHER$'s Exact Swin-B-22k Backbone and Training Schedule, and Are \emph{Single-task} Classifiers Differing Only in the Loss/Sampling Strategy}
\label{tab:lt_baselines}
\resizebox{\columnwidth}{!}{%
\begin{tabular}{l c c c c c}
\toprule
Method & Macro-F1 & Majority & Neutral & Minority & Ultra-Rare \\
\midrule
\multicolumn{6}{l}{\textit{Long-tailed methods (Swin-B backbone, single-task)}} \\
Balanced-Softmax & \colorbox{red!15}{53.36} & 95.97 & 89.07 & 83.82 & 59.50 \\
LDAM-DRW & \colorbox{red!35}{52.61} & \colorbox{blue!15}{96.03} & 89.07 & \colorbox{red!15}{82.93} & \colorbox{red!15}{56.95} \\
cRT (Decoupling) & 54.92 & \colorbox{blue!35}{96.12} & \colorbox{blue!35}{89.66} & \colorbox{blue!15}{84.00} & \colorbox{blue!15}{59.98} \\
\midrule
\multicolumn{6}{l}{\textit{Reference (as reported in \cite{Fishvista})}} \\
Swin-B & \colorbox{blue!15}{55.10} & \colorbox{red!35}{92.6} & \colorbox{red!35}{86.2} & \colorbox{red!35}{79.6} & \colorbox{red!35}{50.4} \\
$\FISHER$ (ours) & \colorbox{blue!35}{60.80} & \colorbox{red!15}{95.5} & \colorbox{blue!15}{89.0} & \colorbox{blue!35}{86.2} & \colorbox{blue!35}{63.80} \\
\bottomrule
\end{tabular}%
}
}
\vspace{-1em}
\end{table}

%------------------------ Table VI -------------------------
\begin{table*}[h]
\centering
\caption{Trait Identification Results on In-species Test Set (highlighted as \colorbox{blue!35}{Best}, \colorbox{blue!15}{Second best}, \colorbox{red!35}{Worst}, \colorbox{red!15}{Second worst})}
\resizebox{0.9\textwidth}{!}{
\begin{tabular}{lccccccccccccc}
\toprule
\multirow{2.5}{*}{{Model}} & \multicolumn{5}{c}{{Average Precision}} & \multicolumn{4}{c}{{F1@0.5}} & \multicolumn{4}{c}{{F1@optimal threshold}} \\ 
\cmidrule(lr){2-6} \cmidrule(lr){7-10} \cmidrule(lr){11-14}
 & mAP & Adip & Pelv & Barb & Dors & Adip & Pelv & Barb & Dors & Adip & Pelv & Barb & Dors \\ \midrule
% --- Representative CNNs ---
ResNext-50 & \csw{91.44} & 98.53 & \csw{74.64} & \csw{97.03} & \csw{95.54} & 96.07 & 86.56 & 96.32 & \csw{94.09} & 96.90 & \csw{85.23} & 96.45 & \csw{94.99} \\
ConvNext-Base & 97.46 & 99.54 & 92.33 & \cs{98.67} & \cs{99.32} & 98.62 & 92.31 & \cb{98.28} & 98.46 & 98.62 & 95.93 & \cb{98.19} & 98.45 \\
\midrule
% --- Representative ViTs ---
ViT-B-16 & \cw{86.95} & \cw{92.93} & \cw{67.70} & \cw{93.29} & \cw{93.89} & \cw{91.51} & \cw{66.75} & \cw{92.41} & \cw{92.57} & \cw{91.71} & \cw{81.23} & \cw{92.14} & \cw{92.68} \\
Swin-B-22k & 95.21 & \csw{98.05} & 86.94 & 97.55 & 98.28 & \csw{95.95} & \csw{86.37} & \csw{96.07} & 96.47 & \csw{96.12} & 91.29 & \csw{96.13} & 96.62 \\
\midrule
% --- SOTA Competitor ---
Q2L (R34-22k SH) & 97.78 & 99.47 & 93.91 & 98.47 & 99.26 & 97.65 & 97.36 & 97.35 & 98.27 & 97.90 & 95.10 & 97.73 & 98.45 \\
Q2L (SwinB-22k SH) & 98.32 & \cb{99.81} & 94.94 & \cb{99.00} & \cb{99.53} & \cb{99.09} & 97.36 & \cs{97.82} & \cb{99.19} & \cs{98.93} & 95.66 & \cs{98.03} & \cb{99.06} \\
\midrule
% --- Proposed Methods ---
MTLSwinB & \cb{99.38} & \cs{99.75} & \cb{99.99} & 98.47 & 99.29 & 98.57 & \cb{99.97} & 97.19 & 98.21 & 98.87 & \cs{99.97} & 97.15 & 98.43 \\
$\FISHER$  & \cs{99.11} & 99.58 & \cs{99.96} & 97.85 & 99.03 & \cs{99.05} & \cb{99.97} & 97.67 & \cs{98.52} & \cb{99.06} & \cb{99.98} & 97.62 & \cs{98.60} \\
\bottomrule 
\end{tabular}
}
\label{tab:in_species}
\vspace{-0.5em}
\end{table*}

%------------------------ Table VII -------------------------
\begin{table*}[!ht]
\centering
\caption{Trait Identification Results on Leave-out-species Test Set (highlighted as \colorbox{blue!35}{Best}, \colorbox{blue!15}{Second best}, \colorbox{red!35}{Worst}, \colorbox{red!15}{Second worst})}
\resizebox{0.9\textwidth}{!}{
\begin{tabular}{lccccccccccccc}
\toprule
\multirow{2.5}{*}{{Model}} & \multicolumn{5}{c}{{Average Precision}} & \multicolumn{4}{c}{{F1@0.5}} & \multicolumn{4}{c}{{F1@optimal threshold}} \\ 
\cmidrule(lr){2-6} \cmidrule(lr){7-10} \cmidrule(lr){11-14}
 & mAP & Adip & Pelv & Barb & Dors & Adip & Pelv & Barb & Dors & Adip & Pelv & Barb & Dors \\ 
 \midrule
% --- Representative CNNs ---
ResNext-50 & \colorbox{red!15}{53.00} & 91.16 & \colorbox{red!35}{1.42} & \colorbox{red!15}{72.86} & 46.56 & 93.28 & \colorbox{red!35}{48.29} & \colorbox{red!15}{81.23} & 73.40 & \colorbox{red!15}{82.29} & \colorbox{red!35}{48.71} & \colorbox{red!15}{80.91} & 72.85 \\
% --- Representative ViTs ---
ViT-B-16 & \colorbox{red!35}{47.27} & \colorbox{red!35}{58.54} & \colorbox{red!15}{40.13} & \colorbox{red!35}{67.60} & \colorbox{red!35}{22.82} & \colorbox{red!35}{69.78} & \colorbox{red!15}{66.66} & \colorbox{red!35}{70.20} & 62.41 & \colorbox{red!35}{69.27} & \colorbox{red!15}{65.89} & \colorbox{red!35}{76.86} & \colorbox{red!15}{61.87} \\
Swin-B-22k & 68.18 & \colorbox{red!15}{89.00} & 59.52 & 80.03 & \colorbox{red!15}{44.18} & 92.04 & 75.96 & 85.03 & \colorbox{red!35}{57.88} & 91.50 & 74.77 & 83.88 & \colorbox{red!35}{58.25} \\
\midrule
% --- SOTA Competitor ---
Q2L (R34-22k SH) & 79.86 & 92.93 & 67.54 & 89.14 & 69.82 & \colorbox{red!15}{88.54} & 85.56 & 86.14 & 70.02 & 92.79 & 88.51 & 90.09 & \colorbox{blue!15}{78.74} \\
Q2L (SWIN-22k SH) & 88.41 & \colorbox{blue!15}{98.61} & \colorbox{blue!15}{93.06} & \colorbox{blue!15}{97.30} & 64.65 & \colorbox{blue!15}{97.84} & 92.75 & \colorbox{blue!15}{96.08} & \colorbox{blue!15}{74.50} & 97.42 & 93.98 & \colorbox{blue!15}{95.22} & 75.43 \\
\midrule
% --- Proposed Methods ---
MTLSwinB & \colorbox{blue!15}{92.71} & 98.41 & \colorbox{blue!35}{99.95} & 95.74 & \colorbox{blue!15}{76.75} & 97.14 & \colorbox{blue!15}{99.90} & 91.60 & \colorbox{red!15}{59.49} & \colorbox{blue!15}{97.87} & \colorbox{blue!35}{99.97} & 85.14 & 72.57 \\
% AquaDeHi & \colorbox{blue!15}{94.12} & \colorbox{blue!35}{99.96} & \colorbox{blue!15}{99.93} & 95.10 & \colorbox{blue!15}{81.46} & \colorbox{blue!35}{98.59} & 99.74 & \colorbox{red!15}{73.74} & \colorbox{blue!15}{75.93} & \colorbox{blue!35}{97.90} & 99.84 & 83.56 & 77.08 \\
$\FISHER$ & \colorbox{blue!35}{97.72} & \colorbox{blue!35}{98.62} & \colorbox{blue!35}{99.95} & \colorbox{blue!35}{99.43} & \colorbox{blue!35}{92.90} & \colorbox{blue!35}{98.59} & \colorbox{blue!35}{99.95} & \colorbox{blue!35}{96.30} & \colorbox{blue!35}{85.45} & \colorbox{blue!35}{97.90} & \colorbox{blue!15}{99.95} & \colorbox{blue!35}{97.06} & \colorbox{blue!35}{89.45} \\
\bottomrule
\end{tabular}%
}
\label{tab:Leave-Out-Species}
\vspace{-0.5em}
\end{table*}

%------------------------ Table VIII -------------------------
\begin{table*}[!ht]
\centering
\caption{Trait Identification Results on Manual-annotation Test Set (highlighted as \colorbox{blue!35}{Best}, \colorbox{blue!15}{Second best}, \colorbox{red!35}{Worst}, \colorbox{red!15}{Second worst})}
\resizebox{0.9\textwidth}{!}{
\begin{tabular}{lccccccccccccc}
\toprule
\multirow{2.5}{*}{{Model}} & \multicolumn{5}{c}{{Average Precision}} & \multicolumn{4}{c}{{F1@0.5}} & \multicolumn{4}{c}{{F1@optimal threshold}} \\ 
\cmidrule(lr){2-6} \cmidrule(lr){7-10} \cmidrule(lr){11-14}
 & mAP & Adip & Pelv & Barb & Dors & Adip & Pelv & Barb & Dors & Adip & Pelv & Barb & Dors \\ 
 \midrule
% --- Representative CNNs ---
ResNext-50 & \colorbox{red!15}{43.25} & 57.02 & \colorbox{red!35}{19.56} & \colorbox{red!35}{36.62} & \colorbox{red!15}{59.79} & 70.12 & \colorbox{red!35}{52.37} & \colorbox{red!35}{63.25} & \colorbox{red!35}{62.53} & 73.58 & \colorbox{red!35}{52.37} & \colorbox{red!35}{62.67} & 68.89 \\
% --- Representative ViTs ---
ViT-B-16 & \colorbox{red!35}{37.63} & \colorbox{red!35}{37.69} & \colorbox{red!15}{26.92} & \colorbox{red!15}{36.72} & \colorbox{red!35}{49.20} & \colorbox{red!15}{64.72} & \colorbox{red!15}{58.76} & 66.35 & \colorbox{red!15}{66.57} & \colorbox{red!35}{65.74} & \colorbox{red!15}{60.47} & 66.61 & 67.57 \\
Swin-B-22k & 51.53 & \colorbox{red!15}{55.55} & 32.20 & 50.54 & 67.82 & 72.51 & 65.99 & \colorbox{blue!15}{73.78} & 75.45 & 72.21 & 63.00 & 71.46 & 74.37 \\ 
\midrule
% --- SOTA Competitor ---
Q2L (R34-22k SH) & 55.84 & 61.67 & 36.80 & 50.44 & \colorbox{blue!35}{74.43} & \colorbox{blue!35}{76.16} & 67.35 & 73.61 & \colorbox{blue!35}{80.07} & \colorbox{blue!15}{75.23} & 69.25 & \colorbox{blue!15}{72.76} & \colorbox{blue!35}{79.63} \\
Q2L (SWIN-22k SH) & 59.39 & 59.18 & 41.65 & 63.46 & \colorbox{blue!15}{73.28} & \colorbox{blue!15}{75.74} & 70.83 & \colorbox{blue!35}{79.16} & \colorbox{blue!15}{78.69} & 74.80 & 70.54 & \colorbox{blue!35}{77.78} & \colorbox{blue!15}{78.44} \\
\midrule
% --- Proposed Methods ---
MTLSwinB & \colorbox{blue!15}{77.34} & \colorbox{blue!15}{74.74} & \colorbox{blue!35}{95.38} & \colorbox{blue!15}{66.85} & 72.38 & \colorbox{red!35}{62.89} & \colorbox{blue!15}{97.14} & \colorbox{red!15}{63.44} & 67.24 & \colorbox{red!15}{66.06} & \colorbox{blue!15}{97.26} & \colorbox{red!15}{63.55} & \colorbox{red!35}{62.59} \\
% AquaDeHi & \colorbox{blue!15}{78.97} & \colorbox{blue!15}{80.26} & \colorbox{blue!15}{95.21} & \colorbox{blue!35}{69.94} & 70.47 & 69.44 & 96.99 & 68.09 & \colorbox{red!35}{62.09} & 71.17 & \colorbox{blue!15}{97.26} & 66.34 & \colorbox{red!35}{58.88} \\
$\FISHER$ & \colorbox{blue!35}{79.53} & \colorbox{blue!35}{83.65} & \colorbox{blue!15}{95.11} & \colorbox{blue!35}{68.87} & 70.50 & 75.38 & \colorbox{blue!35}{97.34} & 70.70 & 67.71 & \colorbox{blue!35}{78.60} & \colorbox{blue!35}{97.29} & 70.97 & \colorbox{red!15}{66.67} \\
\bottomrule
\end{tabular}%
} \vspace{-1em}
\label{tab:manual}
\end{table*}

\vspace{5pt}\noindent\textbf{Implementation details}:
Our framework is implemented in PyTorch. We utilize a Swin-Transformer Base (Swin-B)~\cite{swin} pre-trained on ImageNet-22k as the shared backbone. The input images are resized to $224 \times 224$. We train for $50$ epochs with a batch size of $32$ on a single NVIDIA A5000 GPU using the AdamW optimizer (learning rate $1 \times 10^{-4}$, weight decay $0.01$), including $5$ epochs for the warm-up stage. To manage multi-task optimization, we use homoscedastic uncertainty weighting~\cite{8578879} to dynamically balance $\mathcal{L}_{c}$, $\mathcal{L}_{t}$, and $\mathcal{L}_{s}$ as in Eq. \eqref{totalloss}. The hyperparameter details are summarized in Table \ref{tab:implementation_details}. \ph{The code and trained models are publicly available at \url{https://phucngvinuni.github.io/FISHER/}.}

%------------------------ Table VIII -------------------------

\subsection{Results on Fish-Vista}

\noindent\textbf{Species classification}:
Table~\ref{tab:classification} compares $\FISHER$ with representative baselines from different model families. \ph{The baselines are selected to cover standard CNN and Transformer architectures, large-scale foundation models evaluated under linear probing, and dedicated fine-grained recognition methods. For clarity, the table groups CNNs, Transformers, and foundation models together, while INTR and TransFG represent fine-grained and specialized approaches. This organization enables a broad comparison of $\FISHER$ across different representation-learning paradigms and task-specific methods.} $\FISHER$ achieves the highest Macro-F1 score of $60.8\%$, outperforming the strong Swin-B baseline at $55.1\%$ and specialized FGVC methods such as TransFG~\cite{he2022transfg} at $50.3\%$.

A clear trend emerges across class frequencies. While generic models such as MaxViT \cite{maxvit} and Swin-Transformers \cite{swin} perform well on \textit{Majority} classes ($\sim$94.4\% and $\sim$92.6\%), their performance drops sharply on \textit{ultra-rare} species (\textit{e.g.}, VGG-19 falls to $45.9\%$), indicating a bias toward dominant visual patterns. In contrast, $\FISHER$ improves ultra-rare accuracy by $+13.4\%$ over the Swin-B-22k baseline ($63.8\%$ vs. $50.4\%$). This result supports our negative-transfer hypothesis: by \textit{detaching gradient flow}, the \textit{segmentation} and trait modules learn morphological features that are less biased toward frequent classes. As a result, even species with fewer than $10$ samples benefit from transferable trait cues (\textit{e.g.}, specific \textit{fin shapes}), improving recognition in the long tail. \ph{We further compare three specialized long-tailed methods (Balanced-Softmax~\cite{ren2020balanced}, LDAM-DRW~\cite{cao2019ldam}, and cRT~\cite{kang2020decoupling}) under the same Swin-B backbone and training schedule in Table~\ref{tab:lt_baselines}. Even the strongest of them (cRT) trails $\FISHER$ by $+5.88\%$ Macro-F1, confirming that reweighting the classification head alone cannot match representation-level gradient decoupling.}

\noindent\textbf{Trait identification}: We evaluate trait identification across three levels of difficulty, including In-Species, Leave-Out-Species, and Manual-Annotation, as shown in Tables \ref{tab:in_species}, \ref{tab:Leave-Out-Species}, and \ref{tab:manual}, respectively.

As shown in Table \ref{tab:in_species}, $\FISHER$ performs comparably to the SOTA multi-label transformer Query2Label (Q2L) \cite{liu2021query2label} on the In-Species task, achieving $99.11\%$ mAP versus $97.78\%$ mAP of R34-22k SH, demonstrating strong performance on seen data. In the more challenging Leave-Out-Species setting (Table~\ref{tab:Leave-Out-Species}), which evaluates out-of-distribution (Leave-Out-Species) generalization, conventional methods exhibit substantial degradation (\textit{e.g.}, ViT-B-16 drops to $47.27\%$ mAP). Even the strong baseline Q2L (Swin-22k SH) decreases to $88.41\%$ mAP. In contrast, our method maintains high performance at $97.72\%$ mAP, outperforming Q2L by $9.31\%$. At the level of individual \textit{traits}, it achieves $99.95\%$ AP for \textit{Pelvic fin}, compared to $93.06\%$ from Q2L \cite{liu2021query2label}, indicating that the gradient-decoupled design learns generalized anatomical representations rather than species-specific correlations. On the manually annotated dataset (Table \ref{tab:manual}), the proposed framework remains competitive across \textit{traits}. Notably, for the \textit{Adipose fin}, a small and relatively rare \textit{trait}, it nearly doubles the performance of standard baselines ($83.65\%$ vs. $37.69\%$ for ViT-B-16 \cite{ViT}), demonstrating robustness to image quality variations and domain shifts. Furthermore, it achieves the best overall mAP of $79.53\%$, clearly outperforming the pure classification backbone Swin-B \cite{swin} ($51.53\%$).

%------------------------ Table IX -------------------------
\begin{table*}[]
\small
\centering
\caption{Segmentation Performance Comparison (mIoU in \%) including Input Size.\\Results Are Highlighted as \colorbox{blue!35}{Best}, \colorbox{blue!15}{Second best}, \& \colorbox{red!35}{Worst} (excluding Zero-Shot)}
\setlength\tabcolsep{2.5pt}
\renewcommand{\arraystretch}{1.1}
\resizebox{0.7\textwidth}{!}{
\begin{tabular}{lcccccccccccc}
\toprule
\multirow{2.5}{*}{Model} & \multirow{2.5}{*}{Size} & \multirow{2.5}{*}{mIoU} & \multicolumn{10}{c}{Trait-wise IoU} \\
\cmidrule(lr){4-13}
 & & & \textit{BG} & \textit{Head}  & \textit{Eye} & \textit{Dorsal} & \textit{Pect.} & \textit{Pelvic} & \textit{Anal} & \textit{Caudal} & \textit{Adip.} & \textit{Barbel} \\
    \midrule
    % --- Representative Seg Models ---
    PSPNet \cite{zhao2017pspnet} & \multirow{4}{*}{\textbf{$320\times320$}} & \colorbox{red!35}{73.8} & \colorbox{red!35}{94.6} & \colorbox{red!35}{84.3} & \colorbox{red!35}{77.7} & \colorbox{red!35}{85.1} & \colorbox{red!35}{67.1} & 80.5 & \colorbox{red!35}{83.0} & \colorbox{red!35}{88.7} & \colorbox{red!35}{56.9} & 20.1 \\ 
    DeepLabV3+ \cite{deeplabv3plus2018} &  & 77.0 & 95.4 & 86.0 & 79.1 & 88.1 & 71.0 & \colorbox{blue!35}{84.7} & 86.2 & 89.9 & 66.1 & 23.7 \\ 
    Mask2Former \cite{cheng2021per}&  & \colorbox{blue!15}{81.6} & 95.5 & 86.4 & 79.1 & 87.6 & 74.2 & \colorbox{red!35}{76.1} & 84.7 & 88.8 & 59.6 & \colorbox{red!35}{0.0} \\ 
    YOLOv8 \cite{varghese2024yolov8} &  & \colorbox{blue!35}{83.1} & \colorbox{blue!15}{96.8} & 84.5 & 83.1 & 88.0 & \colorbox{blue!35}{78.0} & 77.5 & 85.6 & 89.6 & 66.8 & 26.7 \\ 
    \midrule
    % --- Our Baseline ---
    MTLSwinB & \textbf{$224\times224$} & 80.4 & 96.7 & 87.8 & 82.7 & 91.5 & 73.5 & 82.9 & 88.5 & \colorbox{blue!15}{92.9} & \colorbox{blue!35}{72.7} & \colorbox{blue!15}{34.5} \\ \midrule
    
    \multirow{2}{*}{$\FISHER$} & \textbf{$224\times224$} & 79.8 & \colorbox{blue!15}{96.8} & \colorbox{blue!15}{88.3} & \colorbox{blue!15}{83.2} & \colorbox{blue!15}{91.6} & 74.0 & 83.0 & \colorbox{blue!15}{89.0} & \colorbox{blue!15}{92.9} & 66.8 & 32.9 \\
      
     & $320\times320$ & 81.4 & \colorbox{blue!35}{97.0} & \colorbox{blue!35}{88.9} & \colorbox{blue!35}{85.8} & \colorbox{blue!35}{92.4} & \colorbox{blue!15}{76.5} & \colorbox{blue!15}{84.3} & \colorbox{blue!35}{89.4} & \colorbox{blue!35}{93.3} & \colorbox{blue!15}{69.7} & \colorbox{blue!35}{36.3} \\
    \midrule
    % --- Zero-shot ---
    MOLMO + SAM \cite{deitke2024molmo} & - & 39.1 & 85.3 & 37.4 & 29.8 & 50.3 & 29.7 & 38.5 & 36.1 & 83.4 & 0.4 & 0.6 \\ 
    \bottomrule
\end{tabular}
}
\label{tab:segmentation}
\vspace{-1.0em}
\end{table*}

%------------------------ Table X -------------------------
\begin{table}[!ht]
\centering
\textcolor{black}{
\setlength{\fboxsep}{2pt}       
\renewcommand{\arraystretch}{1.35} 
\caption{Multi-seed Robustness on Fish-Vista (Mean$\pm$std over $3$ Random Seeds, highlighted as \colorbox{blue!35}{Best} and \colorbox{red!35}{Worst})}
\resizebox{\columnwidth}{!}{%
\begin{tabular}{l c c c c}
\toprule
Metric & FISHER best & \shortstack{$\FISHER$\\(mean$\pm$std)} &
\shortstack{Parallel MTL\\(mean$\pm$std)} & \shortstack{Swin-B\\single-task} \\
\midrule
Species F1 (macro)     & 60.8  & \colorbox{blue!35}{$60.20\pm0.56$} & $58.09\pm0.86$ & \colorbox{red!35}{$54.55\pm0.65$} \\
Ultra-Rare accuracy    & 63.8  & \colorbox{blue!35}{$63.20\pm0.55$} & $60.57\pm0.99$ & \colorbox{red!35}{$49.80\pm0.90$} \\
\midrule
Segmentation mIoU      & 79.8  & \colorbox{blue!35}{$80.01\pm0.31$} & $79.15\pm0.59$ & -- \\
\midrule
Trait mAP (In-Species) & 99.11 & $99.38\pm0.23$  & \colorbox{blue!35}{$99.48\pm0.16$} & -- \\
Trait mAP (Leave-Out-Species)  & 97.72 & \colorbox{blue!35}{$95.27\pm2.16$} & $88.65\pm4.99$ & -- \\
Trait mAP (Manual)     & 79.53 & \colorbox{blue!35}{$80.23\pm0.90$} & $75.71\pm2.11$ & -- \\
\bottomrule
\end{tabular}%
} \vspace{-1em}
\label{tab:seed_robustness}
}
\end{table}

\vspace{3pt}\noindent\textbf{Semantic segmentation}:
Table \ref{tab:segmentation} compares segmentation performance. While $\FISHER$ ($224\times224$) achieves comparable results, we evaluate the $320\times320$ variant for fair comparison with previous works as reported in \cite{Fishvista}. \ph{Segmentation on Fish-Vista is conventionally evaluated at $320\times320$; accordingly, all dense-segmentation baselines are reported at $320\times320$, and we list every method's input size in the ``Size'' column so that $\FISHER$ ($320\times320$) is compared against same-resolution baselines while its $224\times224$ result (the multi-task-native input) is also reported.} The higher-resolution model yields balanced results across biological parts. Although the overall mIoU ($81.4\%$) is slightly lower, likely due to less aggressive background fitting, it achieves $36.3\%$ IoU on \textit{Barbels} and competitive performance on the \textit{Adipose Fin} ($69.7\%$). \ph{This behavior reflects the benefit of attribute-aware prototypes, which maintain cluster centers for each biological part and help preserve small, rare parts often suppressed by gradients from larger regions.} \ph{Furthermore, while the best of three random seeds is reported as the primary result, we also evaluate robustness across all three seeds. As shown in Table~\ref{tab:seed_robustness}, $\FISHER$ achieves $60.20\pm0.56\%$ Macro-F1 and $63.20\pm0.55\%$ Ultra-Rare accuracy, consistently outperforming the Parallel MTL baseline ($58.09\pm0.86\%$ and $60.57\pm0.99\%$, respectively). These results indicate that the gains of $\FISHER$ are consistent across random initializations.}

%------------------------ Table XI -------------------------
% --- Added in revision (R1.3, R3.2): cross-domain validation ---
\begin{table}[!ht]
\centering
\setlength{\fboxsep}{2pt}
\renewcommand{\arraystretch}{1.2}
\caption{Cross-domain Validation on \textbf{CUB-LT} with a $500{:}1$ Imbalance Factor and $96$ Tail Classes (highlighted as \colorbox{blue!35}{Best} and \colorbox{red!35}{Worst}). All Variants Use the Same Swin-B Backbone and Three-head Architecture, Differing Only in Gradient Flow. Results Are Mean$\pm$std over $5$ Seeds}
\label{tab:cub_lt}
\resizebox{\columnwidth}{!}{%
\ph{\begin{tabular}{l c c c}
\toprule
Method & Top-1 (\%) & Tail Acc. (\%) & Attr. mAP (\%) \\
\midrule
Single-Task & $57.96${\scriptsize$\pm1.79$} & $44.57${\scriptsize$\pm1.18$} & -- \\
Parallel MTL (Connected) & \colorbox{red!35}{$57.23${\scriptsize$\pm1.80$}} & \colorbox{red!35}{$42.73${\scriptsize$\pm2.05$}} & $25.24${\scriptsize$\pm0.46$} \\
$\FISHER$ (Detached) & \colorbox{blue!35}{$58.64${\scriptsize$\pm1.98$}} & \colorbox{blue!35}{$45.60${\scriptsize$\pm1.32$}} & \colorbox{blue!35}{$26.88${\scriptsize$\pm0.43$}} \\
\bottomrule
\end{tabular}}%
}
\vspace{-1em}
\end{table}

\subsection{\ph{Cross-Domain Validation on CUB-200-2011}}
\label{sec:cub}
\ph{We evaluate three variants on CUB-200-2011~\cite{wah2011caltech} under matched settings, including \emph{Single-Task}, Parallel MTL with jointly optimized task heads, and $\FISHER$ with boundary-level gradient detachment. As shown in Table~\ref{tab:cub_lt}, Parallel MTL achieves lower mean Top-1 accuracy ($57.23\%$ vs.\ $57.96\%$) and Tail accuracy ($42.73\%$ vs.\ $44.57\%$) than Single-Task, suggesting negative transfer under joint optimization. In contrast, $\FISHER$ achieves $58.64\%$ Top-1 accuracy and $45.60\%$ Tail accuracy, while improving attribute mAP by $1.64$ percentage points over Parallel MTL ($26.88\%$ vs.\ $25.24\%$). Although the classification differences remain within seed variance, the consistent mean ordering Parallel~$<$~Single~$\le$~$\FISHER$, together with the attribute gain, provides additional evidence that hierarchical gradient decoupling can mitigate negative transfer beyond Fish-Vista without increasing model capacity.}

\subsection{Ablation Study}
\label{sec:ablation}

%------------------------ Table XII -------------------------
\begin{table*}[ht]
\centering

\caption{Ablation Study for Comprehensive Structural \& loss. Results Are Highlighted as \colorbox{blue!35}{Best}, \colorbox{blue!15}{Second best}, \colorbox{red!35}{Worst}, \colorbox{red!15}{Second worst}}
\label{tab:structural_ablation_full}

\resizebox{0.93\textwidth}{!}{%
\begin{tabular}{llccccc} 
\toprule
Variant & Configuration / Mechanism & Species F1 & Ultra-Rare Acc. & Leave-Out-Species Trait mAP & Seg mIoU & Barbel IoU \\ 
\midrule

% SECTION A
\multicolumn{7}{l}{\textbf{Scenario 1. }\textit{Component Ablation (Removing Modules)}} \\
Species Only & Species Only & \colorbox{red!15}{55.1} & \colorbox{red!35}{50.4} & -- & -- & -- \\
No Segmentation & Trait $\to$ Species & \colorbox{red!35}{54.7} & \colorbox{red!15}{57.7} & 94.2 & -- & -- \\
No Trait & Seg $\to$ Species & 57.1 & 60.0 & -- & 78.5 & \colorbox{red!15}{27.8} \\
\midrule

% SECTION B
\multicolumn{7}{l}{\textbf{Scenario 2. }\textit{Order \& Connectivity (Changing Topology)}} \\
Standard MTL & Parallel (Shared Backbone) & 57.8 & 60.4 & 92.7 & \colorbox{blue!35}{80.4} & \colorbox{blue!35}{34.5} \\
Reverse Order & Species $\to$ Trait $\to$ Seg ($\Downarrow$) & 58.5 & \colorbox{blue!15}{62.8} & 91.5 & \colorbox{blue!15}{80.1} & \colorbox{blue!15}{33.4} \\
Fully Connected & Seg $\to$ Trait $\to$ Species ($\Uparrow$) & \colorbox{blue!15}{60.1} & 61.4 & 94.1 & \colorbox{red!15}{78.3} & 28.9 \\
\midrule

% SECTION C
\multicolumn{7}{l}{\textbf{Scenario 3. }\textit{Partial Detachment (Negative Transfer)}} \\
Partial A (Seg Connect) & Seg $\to$ Trait $\parallel$ Species & 59.8 & 60.1 & \colorbox{red!35}{86.4} & 79.0 & 31.3 \\
Partial B (Trait Connect) & Seg $\parallel$ Trait $\to$ Species & 59.9 & 61.1 & \colorbox{red!15}{87.4} & 79.5 & 30.7 \\
\midrule

% SECTION D
\multicolumn{7}{l}{\textbf{Scenario 4. }\textit{Loss \& Weighting (Detached Architecture)}} \\
Base Detached & w/o UW and Ortho. Losses & 58.1 & 61.5 & \colorbox{blue!15}{95.7} & 79.2 & 32.6 \\
+ Uncertainty & w/ UW Only & 55.7 & 59.6 & 95.4 & \colorbox{red!35}{77.7} & \colorbox{red!35}{27.3} \\
+ OrthoLoss & w/ Orthogonal Loss Only & 58.1 & 62.2 & 94.5 & 79.1 & 31.9 \\
\midrule

% FINAL ROW
\rowcolor{blue!5} 
$\FISHER$ (Ours) & Detached + UW + Ortho. Losses & \colorbox{blue!35}{60.8} & \colorbox{blue!35}{63.8} & \colorbox{blue!35}{97.7} & 79.8 & 32.9 \\ 
\bottomrule
\end{tabular}
} \vspace{-1.0em}
\end{table*}

We conduct four variant scenarios to analyze the contribution of each component in the proposed framework. Results are summarized in Table \ref{tab:structural_ablation_full}, providing empirical evidence of negative transfer in hierarchical fine-grained recognition.

We first evaluate the role of the three tasks in \textbf{Scenario 1}. The single-task baseline (species only) performs poorly under long-tailed distributions, achieving only $50.4\%$ accuracy on \textit{ultra-rare} species. Removing intermediate semantic stages (\textit{i.e.}, ``no \textit{segmentation}'' or ``no \textit{trait}'') also leads to suboptimal performance. This confirms that explicitly modeling the bottom-up biological hierarchy (\textit{segmentation} $\to$ \textit{traits} $\to$ \textit{species}) is essential for recognizing rare classes with limited training samples.

\ph{To justify the proposed ordering, \textbf{Scenarios~2} and~\textbf{3} together evaluate five hierarchical/connectivity variants: \textit{Standard MTL (Parallel)}, \textit{Reverse Order} (species$\to$trait$\to$seg), \textit{Fully Connected} (bidirectional gradient flow), and the two \textit{Partial Detachment} variants (\textit{Partial A} and \textit{Partial B}), all would be compared against the fully detached $\FISHER$.} In \textbf{Scenario 2}, the fully connected variant allows gradients to propagate from the species head to the \textit{segmentation} module. While it achieves a competitive Species F1 score ($60.1\%$), it sharply degrades \textit{segmentation} quality, with \textit{Barbel} IoU dropping to $28.9\%$ (vs. $33.4\%$ in the detached model). This indicates that without gradient isolation, dominant high-level \textit{classification} objectives suppress fine-grained morphological learning.

\begin{figure}[]
\centering
\includegraphics[width=\columnwidth]{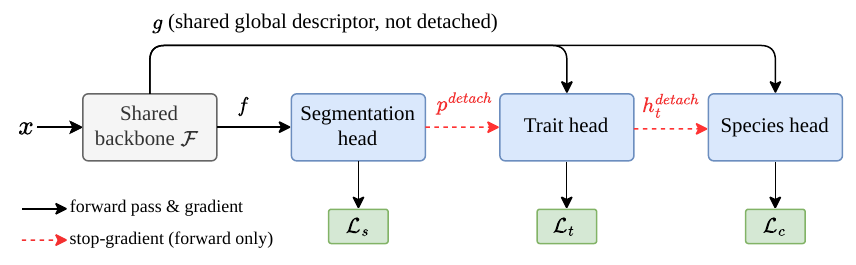}
\caption{\ph{Gradient flow in $\FISHER$. Detached connections block backward gradients at task boundaries while preserving forward information flow}.}
\label{fig:variants}
\end{figure}

The partial detachment settings in \textbf{Scenario 3} further show that allowing gradient flow between any pair of heads harms generalization, reducing Leave-Out-Species \textit{trait} mAP to approximately $86–87\%$. \ph{Each variant is obtained by removing a single stop-gradient from the fully detached model. \textit{Partial A} (Seg~Connect) reconnects the segmentation--trait interface, allowing trait gradients to reach the segmentation branch (Seg~$\to$~Trait), while the trait--species interface remains detached. \textit{Partial B} (Trait~Connect) reconnects the trait--species interface (Trait~$\to$~Species), while the segmentation--trait interface remains detached. The full $\FISHER{}$ detaches \textit{both} interfaces, as illustrated in Fig.~\ref{fig:variants}.} In contrast, the fully detached design maintains Leave-Out-Species mAP above $95\%$, demonstrating that gradient isolation is critical for learning generalized, species-agnostic attribute representations.

\ph{\textbf{Scenario 4} highlights the interaction between loss mechanisms. Applying uncertainty weighting alone to the detached architecture unexpectedly reduces performance (\textit{species} F1 drops from $58.1\%$ to $55.7\%$), suggesting instability or convergence to suboptimal solutions. However, when combined with orthogonality regularization, performance improves substantially, achieving the best results across all metrics (\textit{species} F1: $60.8\%$, \textit{ultra-rare} Acc: $63.8\%$). This indicates a strong synergy: orthogonality regularizes the feature space by enforcing separation, which stabilizes uncertainty weighting for effective task balancing. We attribute the limited benefit of uncertainty weighting alone to the representation geometry on which it operates. With correlated prototypes, adaptive task weighting cannot directly resolve representation-level redundancy and may converge to a suboptimal balance. Orthogonality promotes better-separated prototypes, providing a more structured representation for adaptive weighting. The two mechanisms are therefore complementary: orthogonality regularizes the representation space, while uncertainty weighting balances task contributions. Their combination in $\FISHER$ yields the best overall performance, particularly for ultra-rare species recognition while maintaining high segmentation quality.}

\begin{figure}[]
\centering
\includegraphics[width=\linewidth]{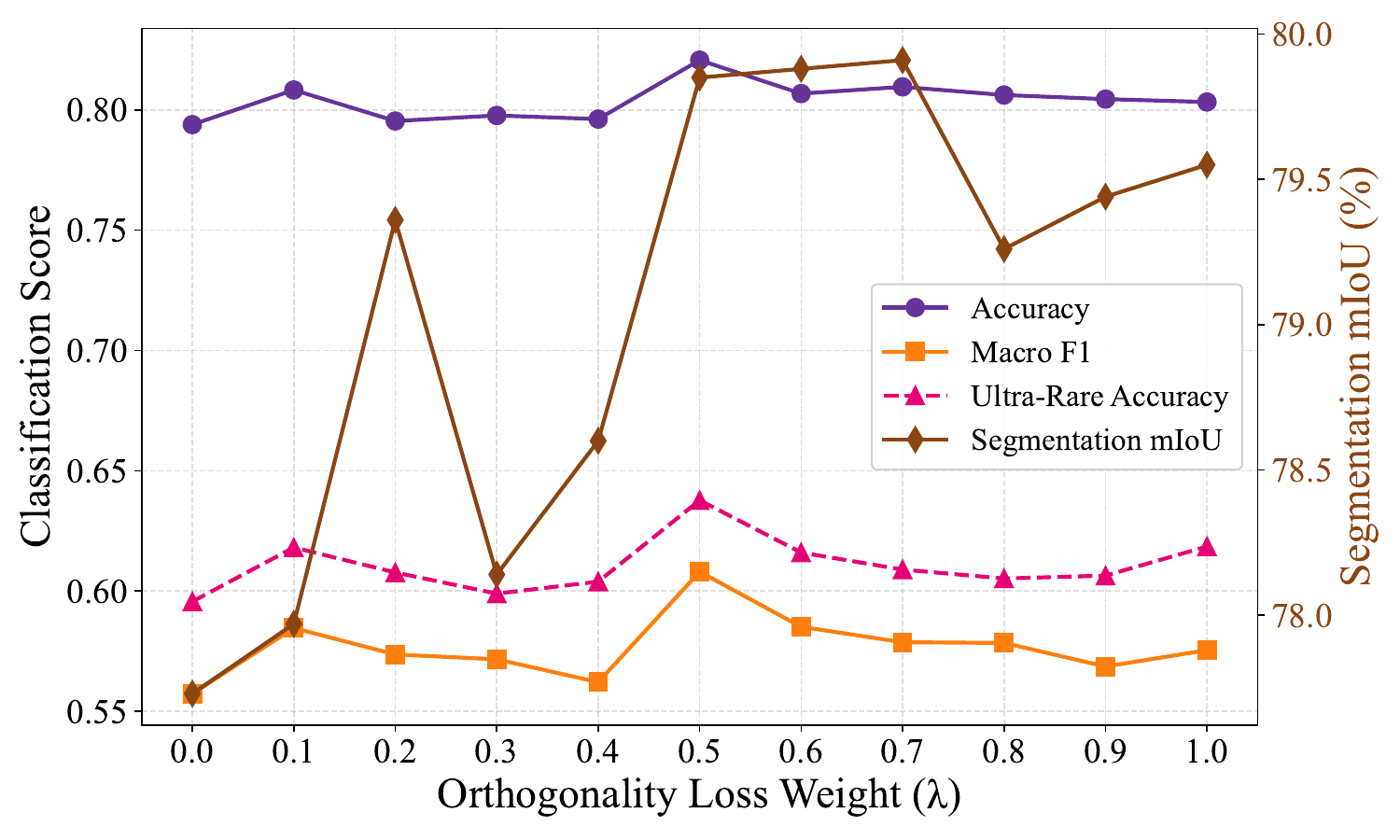}
\vspace{-2.0em}
\caption{Sensitivity analysis of orthogonality weight ($\lambda_{ortho}$) in Eq. (\ref{totalloss}).}
\label{fig:ortho_combined}
\vspace{-0.5em}
\end{figure}

\subsection{Hyperparameter Sensitivity Analysis}
\label{subsec:hyperparam}

\ph{The orthogonality loss weight ($\lambda_{ortho}$) controls the separation between prototype vectors in the attribute-aware head, preventing feature collapse where distinct anatomical parts overlap in the latent space.} We analyze its effect by sweeping $\lambda_{ortho} \in [0.0, 1.0]$. Fig.~\ref{fig:ortho_combined} shows classification and segmentation metrics on a dual-axis plot. While overall accuracy (solid \textcolor{darkpurple}{purple line} with \textcolor{darkpurple}{$\bullet$} marker) remains stable across $\lambda_{ortho}$, a clear relationship emerges between segmentation quality (\textcolor{Mahogany}{brown line} with \textcolor{Mahogany}{$\blacklozenge$} marker) and \textit{ultra-rare} species classification (dashed \textcolor{deeppink}{pink line} with \textcolor{deeppink}{$\blacktriangle$} marker).

When orthogonality is low ($\lambda_{ortho}<0.3$), the model struggles to distinguish visually similar parts, yielding a mean IoU of only $\sim 78\%$ and lower accuracy for rare species. In contrast, overly large values ($\lambda_{ortho} > 0.7$) introduce excessive regularization, constraining the feature space and reducing the model’s ability to capture intra-class variations. We observe a sharp performance transition peaking at $\lambda_{ortho} = 0.5$, where mIoU reaches nearly $80\%$ and aligns with the maximum \textit{ultra-rare} accuracy ($\sim 64\%$). This strong correlation supports our detached hierarchical hypothesis: enforcing well-separated morphological representations directly improves rare-species recognition. Based on this analysis, we select $\lambda_{ortho} = 0.5$ as the final setting, achieving an effective balance between prototype separation and representational flexibility for robust generalization.

\begin{figure}[]
    \centering
    \includegraphics[width=\linewidth]{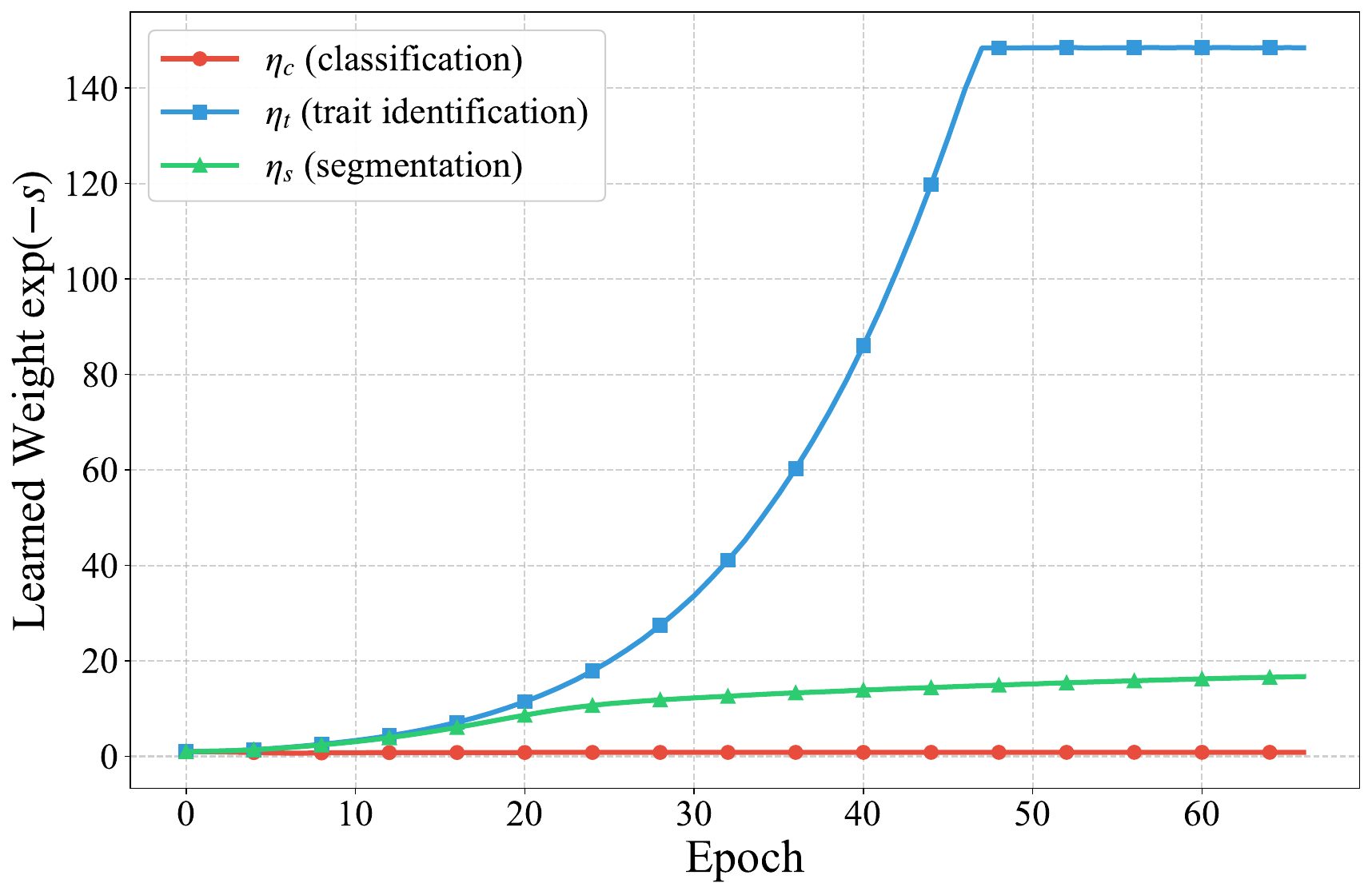}
    \caption{Evolution of the dynamic task weights defined in Eq.~(\ref{uw_loss}), showing the learnable weights ($1/2\sigma^2$) for each task across 50 training epochs.}
    \vspace{-2pt}
    \label{fig:task_weight_evolution}
\end{figure}

\begin{figure}[t]
    \centering
    \includegraphics[width=\linewidth]{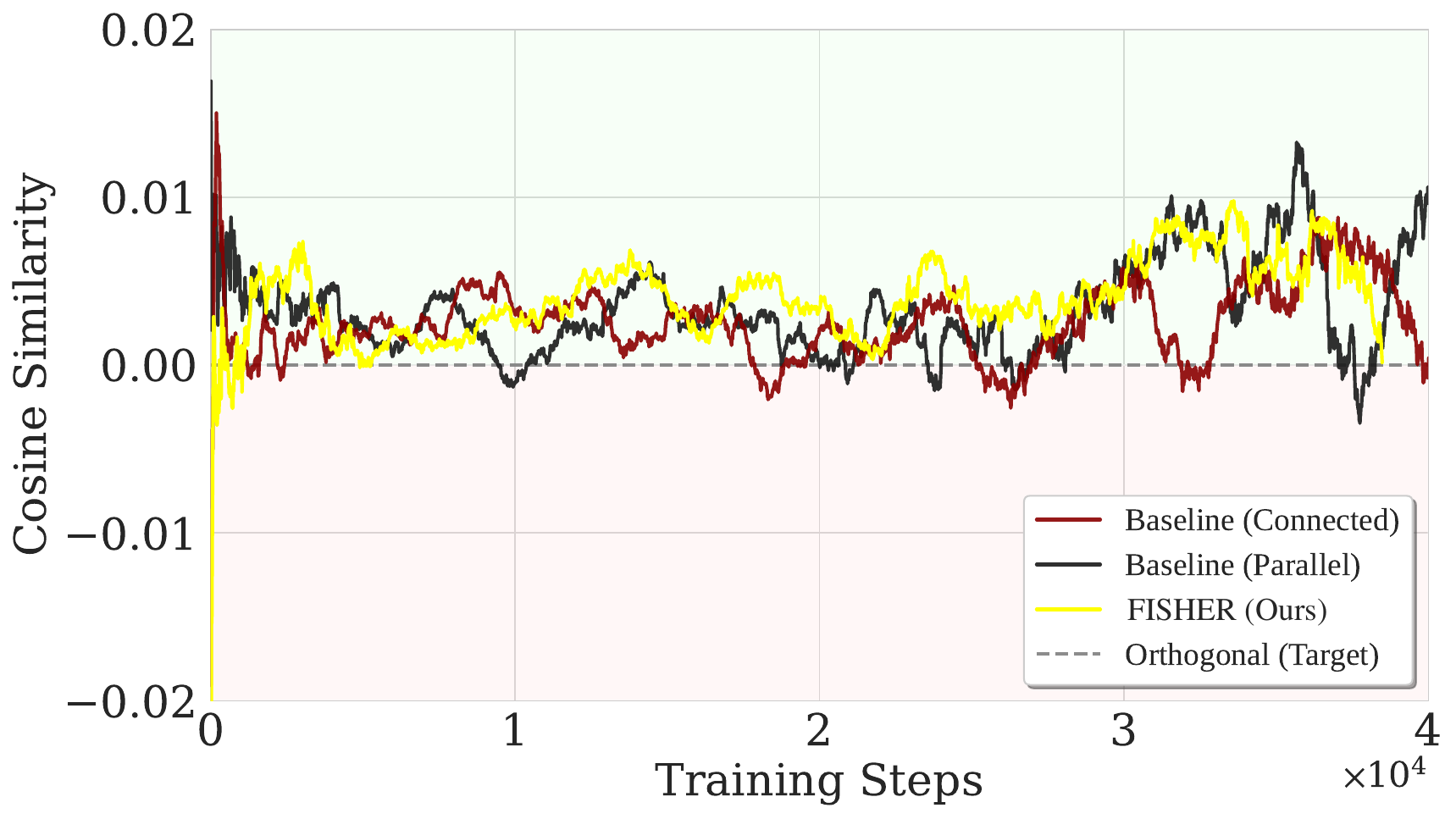}
    \caption{Gradient Alignment in Shared Backbone. Cosine similarity between classification and segmentation gradients: the baseline (black, \textcolor{wine}{dark red}) shows frequent conflicts (negative values), while $\FISHER$ (\textcolor{yellow!95!orange}{yellow}) maintains stable positive alignment.}
    \label{fig:grad_similarity}
    \vspace{-1em}
\end{figure}

%------------------------ Table XIII -------------------------
\begin{table*}[]
\centering
\caption{Ablation to verify the effectiveness of our architectural detachment for multi-task learning. \textbf{Ultra-R} denotes accuracy on the hardest Ultra-Rare species subset ($<10$ samples). All experimental results are extracted by using the Swin-B backbone. GradNorm was re-tuned across learning-rate and $\alpha$ settings; the reported figures use the best configuration.}
\label{tab:gradient_conflict}
\resizebox{0.9\textwidth}{!}{%
\begin{tabular}{l l c c c c}
\toprule
Method & Optimization Strategy & Species F1 (\%) & Ultra-R Acc. (\%) & Trait Leave-Out-Species mAP (\%) & Seg mIoU (\%) \\
\midrule
Standard MTL & Naive Sum ($\sum \mathcal{L}_i$) & 53.8 & 56.1 & 88.8 & 76.9 \\
MTL + UW \cite{8578879} & Uncertainty Weighting & 57.8 & 60.4 & 92.7 & \textbf{80.4} \\
MTL + GradNorm \cite{pmlr-v80-chen18a} & Gradient Normalization & 56.9 & 58.1 & 79.7 & 75.6 \\
MTL + PCGrad \cite{yu2020gradient} & Gradient Projection & 58.1 & 59.6 & 80.2 & 78.3 \\
\midrule
$\FISHER$ & Gradient Detachment & \textbf{60.8} & \textbf{63.8} & \textbf{97.7} & 79.8 \\
\bottomrule
\end{tabular}
}
\end{table*}

\begin{figure*}[t]
    \centering
    \includegraphics[width=\linewidth]{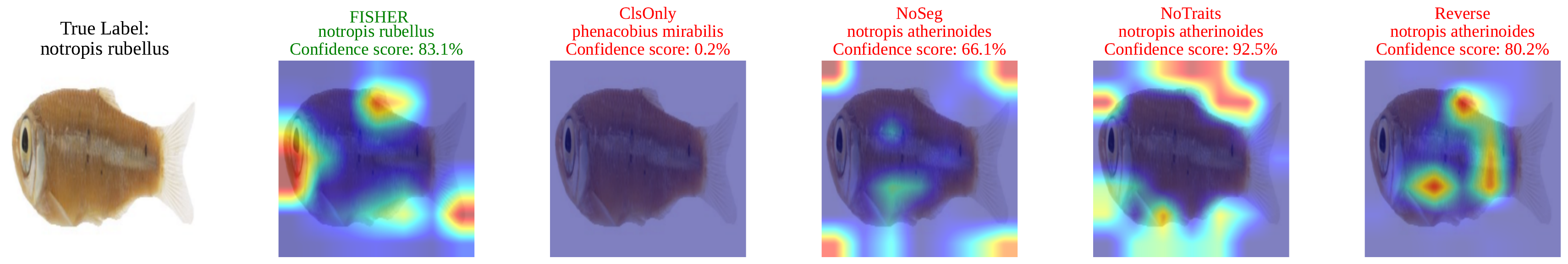}
    \vspace{-15pt} 
    
\caption{Grad-CAM visualization results on a \textit{notropis rubellus} sample. 
From left to right: Original Image, $\FISHER$ (Ours), and ablation baselines: \textit{ClsOnly} (classification only), \textit{NoSeg} (without segmentation branch), \textit{NoTraits} (without attribute-aware prototypes), and \textit{Reverse} (reversed task hierarchy).}
\label{fig:gradcam_vertical}
\vspace{-0.5em}
\end{figure*}

\begin{figure}[]
    \centering
    \includegraphics[width=\linewidth]{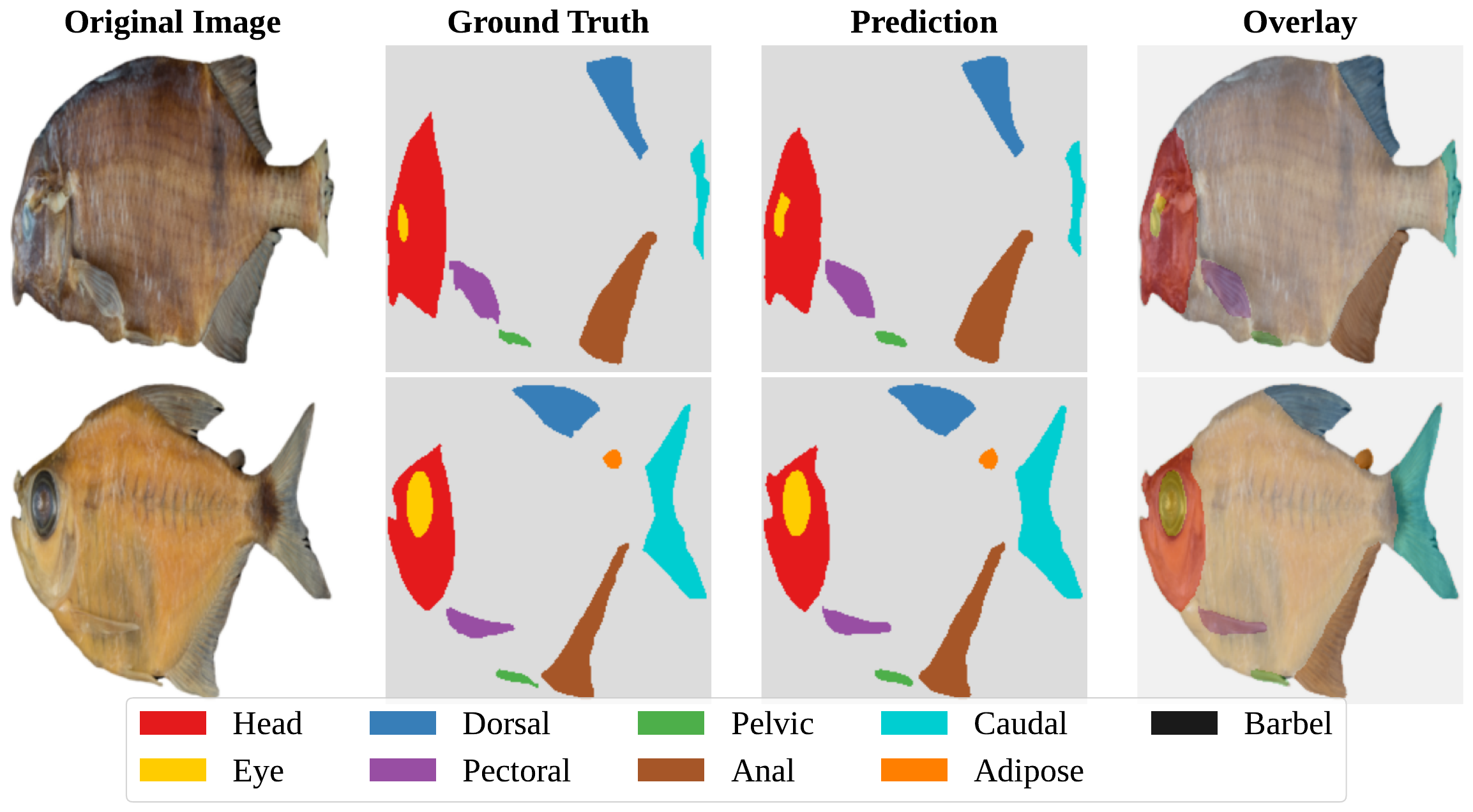}
    \vspace{-1.0em}
    \caption{Visualization of \textit{segmentation} masks generated by $\FISHER$.}
    \label{fig:seg}
    \vspace{-1.0em}
\end{figure}

\subsection{Analysis of Dynamic Task Weighting}
\label{sec:dynamic_weight_analysis}

A key component of $\FISHER$ is the homoscedastic uncertainty weighting strategy, which dynamically balances the losses of the three tasks based on their aleatoric uncertainty. Fig.~\ref{fig:task_weight_evolution} illustrates the evolution of the learnable weights ($1/2\sigma_m^2$) during training. Notably, $\FISHER$ assigns the lowest weight to the species loss (\textcolor{red}{red line}, $<1\%$ relative importance), indicating that species supervision is treated as a high-variance signal. This aligns with the long-tailed nature of the Fish-Vista dataset, where gradients from rare classes are sparse and unstable, and helps prevent noisy signals from dominating the shared backbone, particularly in early training.

In contrast, the weight for \textit{trait identification} (\textcolor{blue}{blue line}) increases rapidly and dominates the total loss ($\sim90\%$ relative importance), suggesting that binary \textit{trait} attributes provide a more stable learning signal with lower uncertainty. The \textit{segmentation} weight (\textcolor{green}{green line}) remains moderate, acting as a bridge between pixel-level details and semantic attributes. This weighting pattern reinforces the proposed detached hierarchical design: with gradient detachment, the backbone learns bottom-up morphological cues from \textit{segmentation} and \textit{trait} supervision, while the species head performs \textit{classification} without destabilizing these representations.

This behavior is further supported by the ablation study in Table \ref{tab:structural_ablation_full}. Removing uncertainty weighting (\textit{i.e.}, using fixed equal weights) causes the species loss to dominate optimization, leading to reduced \textit{ultra-rare} accuracy ($63.8\%\rightarrow61.5\%$) and slightly degraded \textit{segmentation} quality (\textit{Barbel} IoU: $32.9\%\rightarrow32.6\%$). These results highlight that dynamically prioritizing more stable tasks effectively mitigates noise in fine-grained, long-tailed biological data.

\subsection{Gradient Alignment Analysis}
\label{sec:grad_analysis}

To examine negative transfer in the shared Swin-Transformer backbone, we analyze gradient dynamics by computing the cosine similarity ($\mathcal{S}_{cos}$) between the gradients of the classification and segmentation tasks:
\begin{equation}
\mathcal{S}_{cos}=\frac{\nabla_{\theta}\mathcal{L}_{c}\cdot\nabla_{\theta}\mathcal{L}_{s}}{\|\nabla_{\theta}\mathcal{L}_{c}\|\|\nabla_{\theta}\mathcal{L}_{s}\|}
\end{equation}
where $\theta$ denotes the parameters at the final layer of the backbone. A negative value ($\mathcal{S}_{cos} < 0$) indicates gradient conflict, where tasks update parameters in opposing directions.

To further examine how the tasks interact during joint optimization, Fig.~\ref{fig:grad_similarity} visualizes the moving average of gradient cosine similarity, $\mathcal{S}_{cos}$. We compare two representative shared-learning settings, namely \textit{Baseline (Parallel)} with independent task heads and \textit{Baseline (Connected)}, corresponding to $\FISHER$ without gradient detachment. Both baselines exhibit frequent and pronounced oscillations into the negative region, indicating persistent gradient conflicts. These observations suggest that, without explicit gradient control, supervision from long-tailed species classification can interfere with the morphological representations learned by the \textit{segmentation} task.
In contrast, $\FISHER$ maintains a more stable and predominantly positive gradient alignment ($\mathcal{S}_{cos}>0$). \ph{Gradient decoupling at task boundaries prevents conflicting \textit{classification} gradients from directly modifying intermediate \textit{trait} and \textit{segmentation} representations. Specifically, stop-gradients allow classification supervision to reach the shared backbone through the global descriptor $\boldsymbol{g}$ while protecting lower-level task-specific representations from direct higher-level interference. This preserves joint backbone optimization while maintaining stable gradient alignment across the hierarchy.} Consequently, $\FISHER$ reduces the influence of conflicting updates from rare classes while preserving structural morphological cues, leading to more robust and transferable feature representations.

Overall, this analysis highlights that hierarchical gradient decoupling plays a critical role in stabilizing optimization and preserving meaningful morphological representations in fine-grained multi-task learning.

\begin{figure*}
    \centering
    \includegraphics[width=\linewidth]{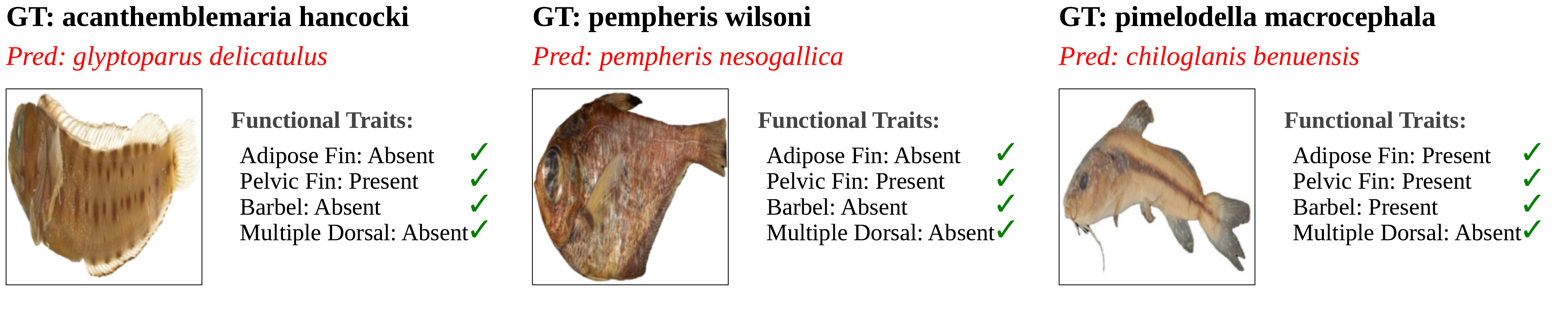}
    \vspace{-2.0em}
    \caption{Visualization results of \textit{trait identification} extracted by the proposed $\FISHER$.}
    \label{fig:trait}
\end{figure*}

\begin{figure*}[h!]
    \centering
    \includegraphics[width=\linewidth]{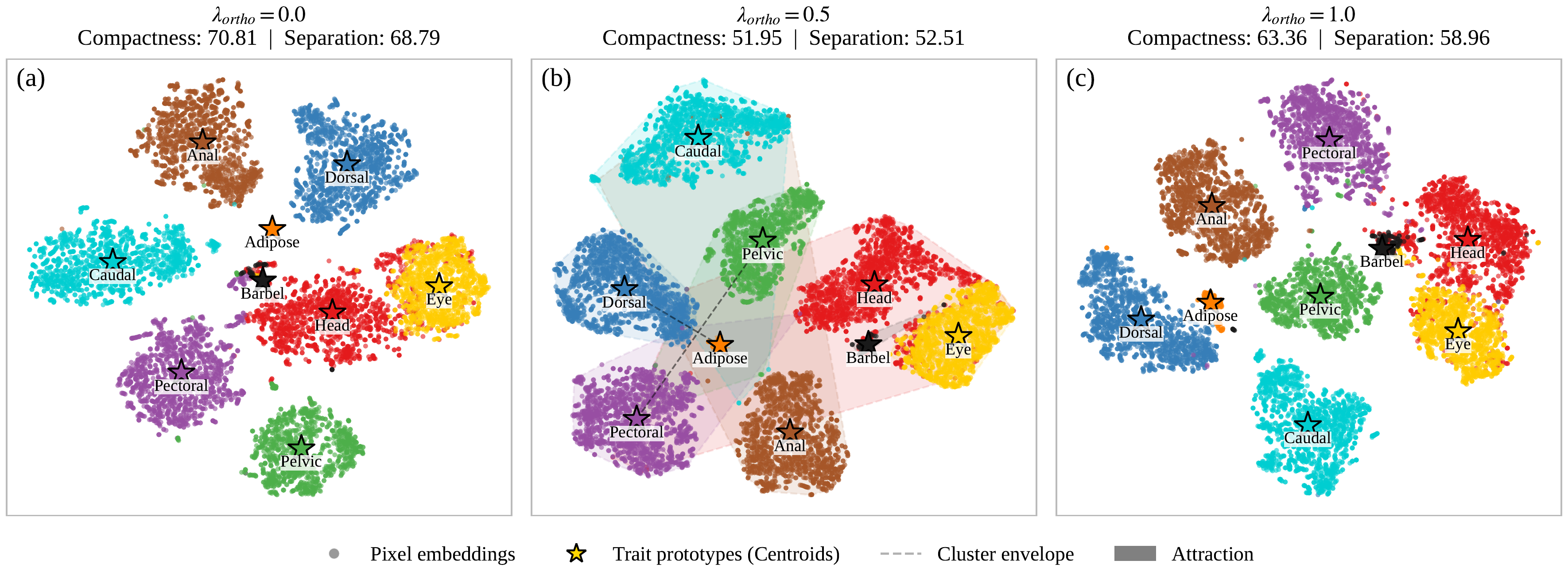}
    \vspace{-2.0em}
    \caption{\ph{t-SNE visualization of the semantic features to illustrate the effectiveness of the orthogonality constraint for \textit{part (anatomical class) identification}.} ($\star$) denotes the learned prototype vectors.
    \textbf{(a)} without orthogonality ($\lambda_{ortho}=0.0$),
    \textbf{(b)} moderate orthogonality ($\lambda_{ortho}=0.5$), 
    and \textbf{(c)} strong orthogonality ($\lambda_{ortho}=1.0$).}
    \label{fig:tsne}
\end{figure*}

\subsection{Task Balancing Strategy}
We benchmark $\FISHER$ against state-of-the-art gradient conflict mitigation methods to evaluate the effectiveness of the proposed architecture. As shown in Table~\ref{tab:gradient_conflict}, general-purpose optimization strategies struggle to capture the semantic hierarchy of biological traits. PCGrad \cite{yu2020gradient}, which resolves conflicting gradients via projection, achieves a competitive F1-score (58.1\%) but performs markedly worse on Leave-Out-Species trait identification (80.2\% mAP vs. 97.7\%). This suggests that enforcing gradient agreement across tasks may suppress task-specific morphological cues critical for generalization. Similarly, uncertainty weighting \cite{8578879} achieves the highest segmentation mIoU (80.4\%) but underperforms on ultra-rare species classification (60.4\% vs. 63.8\%). This supports the negative transfer hypothesis, indicating that optimizing dominant, pixel-dense tasks does not necessarily benefit rare-species recognition. In contrast, by decoupling gradients across hierarchical tasks, $\FISHER$ slightly reduces segmentation precision (0.6\% mIoU) while substantially improving generalization to rare and unseen species.

\subsection{Qualitative Analysis}

\noindent\textbf{Attention analysis}:
To better understand the learning behavior of $\FISHER$ and its ablation variants, we visualize Grad-CAM \cite{Grad_CAM} activation maps in Fig.~\ref{fig:gradcam_vertical}, where \textcolor{red}{red} regions indicate close attention.
\begin{itemize}
\item \textbf{Precise localization of biological traits.}
In the \textit{Notropis rubellus} example, $\FISHER$ demonstrates strong spatial awareness by focusing on key morphological regions, particularly the snout/head area and the bases of the dorsal and anal fins. These regions are biologically diagnostic for the \textit{Notropis} genus. By concentrating on these traits rather than the overall body contour, the model achieves a correct prediction with high confidence (83.1\%), indicating that the trait identification head effectively guides the classification backbone toward taxonomically relevant features.

\item \textbf{Background bias without spatial supervision.} Without explicit spatial guidance, the NoSeg (no \textit{segmentation} branch) baseline fails to attend to the foreground and instead activates strongly near image corners. Similarly, the NoTraits (no \textit{trait identification} head) variant produces fragmented attention scattered across edges and background regions. As a result, both variants incorrectly predict \textit{Notropis atherinoides}.

\item \textbf{Loss of spatial consistency.} The ClsOnly (\textit{classification} only) baseline exhibits nearly uniform activation with extremely low confidence (0.2\%), indicating that classification supervision alone provides insufficient spatial guidance. The reverse variant also produces scattered and misaligned attention along the fish flank, failing to capture key diagnostic regions. These observations suggest that high-level labels alone are insufficient to guide reliable low-level feature localization.
\end{itemize}

\noindent\textbf{Segmentation quality and trait identification}:
As illustrated in Fig. \ref{fig:seg}, $\FISHER$ demonstrates a superior ability to localize fine-grained anatomical structures. Notably, it accurately delineates small and thin parts such as barbels and effectively distinguishes visually similar fins (\textit{e.g.}, \textit{Adipose} vs. \textit{Dorsal}), which are often missed or merged by generic baselines. In Fig. \ref{fig:trait}, we visualize the trait identification results on the Manual Test Set, which contains species not observed during training. While classification performance degrades under this setting, the model remains robust in identifying traits of unseen species, highlighting its strong generalization capability. \ph{Beyond the qualitative visualizations, we provide quantitative evidence of anatomical alignment and cross-task consistency in $\FISHER$. At the part level, the segmentation IoU against expert-annotated masks in Table~\ref{tab:segmentation} confirms that the predicted regions align with anatomical structures, including a \textit{Barbel} IoU of $36.3\%$. At the cross-task level, Table~\ref{tab:consistency} shows high consistency between species and trait predictions ($99.7$--$99.99\%$) and between segmented parts and trait predictions ($93.7$--$97.1\%$), supporting the coherence of the segmentation$\to$trait$\to$species hierarchy. Finally, Table~\ref{tab:seed_robustness} reports a trait mAP of $95.27\%$ on unseen species, indicating that the learned morphological representations remain transferable beyond the species observed during training.}

%------------------------ Table XIV -------------------------
\begin{table}[]
\centering
\textcolor{black}{
\caption{Internal consistency of $\FISHER$'s hierarchical predictions on the
classification test set. \emph{Species$\to$Trait}: fraction of
samples where the predicted species agrees with the predicted trait; \emph{Seg$\to$Trait}:
fraction where the segmented anatomical part agrees with the predicted trait}
\label{tab:consistency}
\small
\renewcommand{\arraystretch}{1.2}
\setlength{\tabcolsep}{8pt}
\begin{tabular}{lcc}
\toprule
Trait & Species$\to$Trait (\%) & Seg$\to$Trait (\%) \\
\midrule
Adipose Fin & 99.86 & 97.06 \\
Pelvic Fin  & 99.99 & 95.10 \\
Barbel      & 99.67 & 93.73 \\
\bottomrule
\end{tabular}
}
\end{table}

\begin{figure}[]
    \centering
    \includegraphics[width=\linewidth]{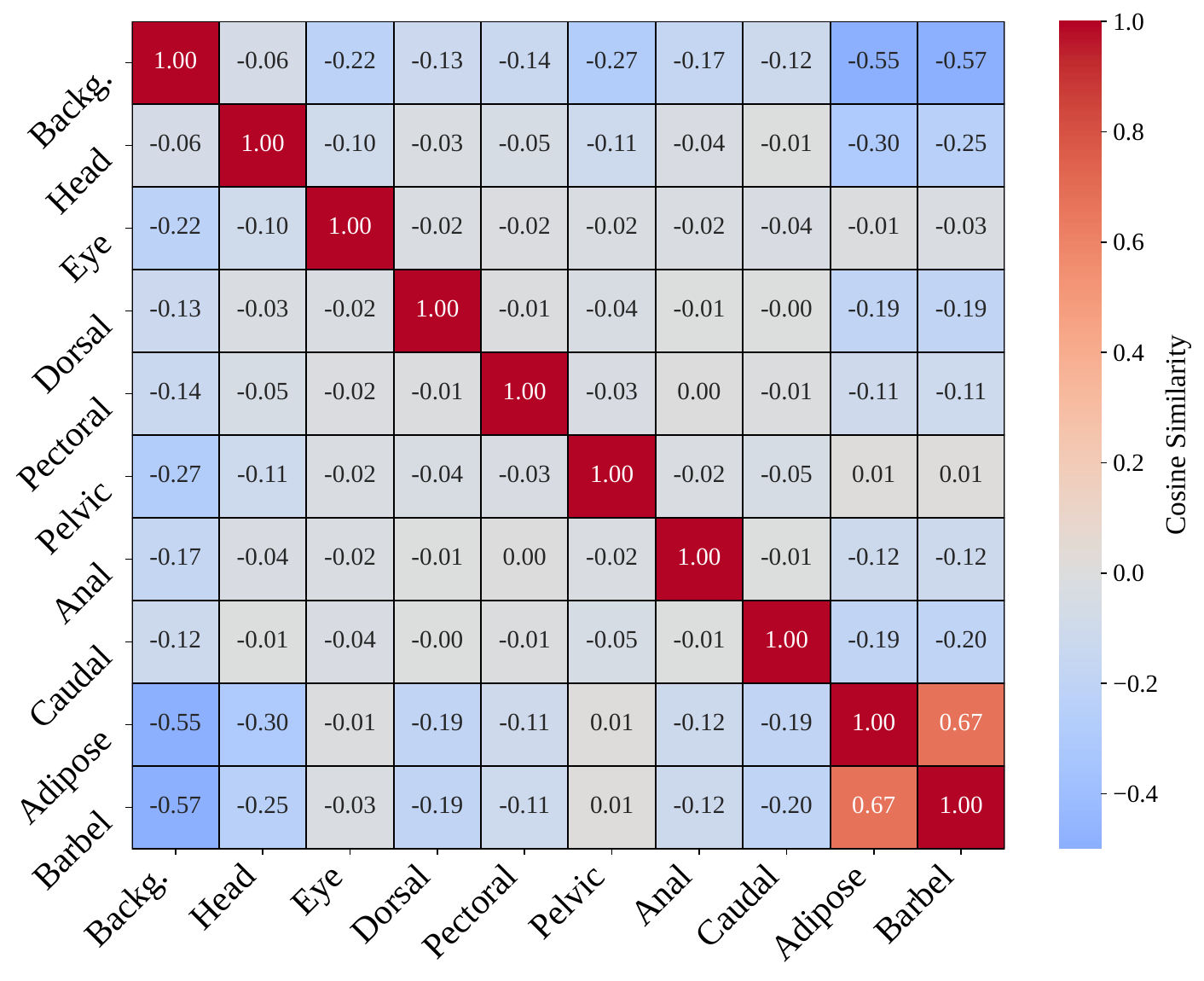}
    \caption{Prototype correlation matrix ($\lambda_{ortho}=0.5$). 
    The heatmap shows the cosine similarity between learned prototype vectors.}
    \label{fig:prototype_corr}
\end{figure}

\subsection{Latent Space Analysis}
We investigate the effect of the orthogonality constraint on the learned feature representations through t-SNE visualization and prototype correlation analysis.

\vspace{3pt}\noindent\textbf{Effect of orthogonality on feature separation}:
\ph{We use t-SNE \cite{tsne} to project extracted representations in the feature space as presented in Fig.~\ref{fig:tsne} for analyzing different parts under varying orthogonality weights ($\lambda_{ortho}$).}

\begin{itemize}

\item At $\lambda_{ortho}=0.0$ (Fig.~\ref{fig:tsne}a), the embeddings form loosely structured clusters in the latent feature space, with considerable overlap between semantically similar parts (\textit{e.g.}, \textit{Dorsal} and \textit{Adipose Fins}). Such poor separability in the embedding space leads to ambiguous representations and reduces segmentation precision.

\item At $\lambda_{ortho}=0.5$ (Fig.~\ref{fig:tsne}b), the embeddings form more compact and well-separated clusters in the latent feature space. The prototype vectors ({\ding{73}} markers) align with the centers of their corresponding semantic clusters, encouraging intra-class cohesion while improving inter-class separability.

\item At $\lambda_{ortho}=1.0$ (Fig.~\ref{fig:tsne}c), the embeddings form highly compact clusters with strong inter-class separation in the latent space. However, this overly rigid structure (Separation Score: $58.96$) may reduce the model's ability to capture intra-class variations in unseen species, leading to a slight drop in Leave-Out-Species performance. In practice, $\lambda_{ortho}=0.5$ provides the best trade-off between discriminability and generalization.

\end{itemize}

\noindent\textbf{Prototype correlation analysis}:
Fig. \ref{fig:prototype_corr} displays the cosine similarity matrix between learned prototypes at $\lambda_{ortho}=0.5$. Ideally, prototypes representing distinct anatomical parts should be orthogonal (correlation $\approx 0$).
The heatmap shows low correlation values ($<0.2$) for most pairs, confirming that $\FISHER$ successfully learns distinct representations for different body parts. Notably, even for spatially adjacent and visually similar parts such as the \textit{Anal} and \textit{Pelvic Fins}, the correlation remains low, highlighting the effectiveness of the proposed attribute-aware learning objective. A slight positive correlation exists between \textit{Adipose} and \textit{Barbel} (0.67), likely due to their shared property as small, high-frequency texture features, yet they remain sufficiently distinct for accurate segmentation.

\begin{figure}[]
    \centering
    \includegraphics[width=\linewidth]{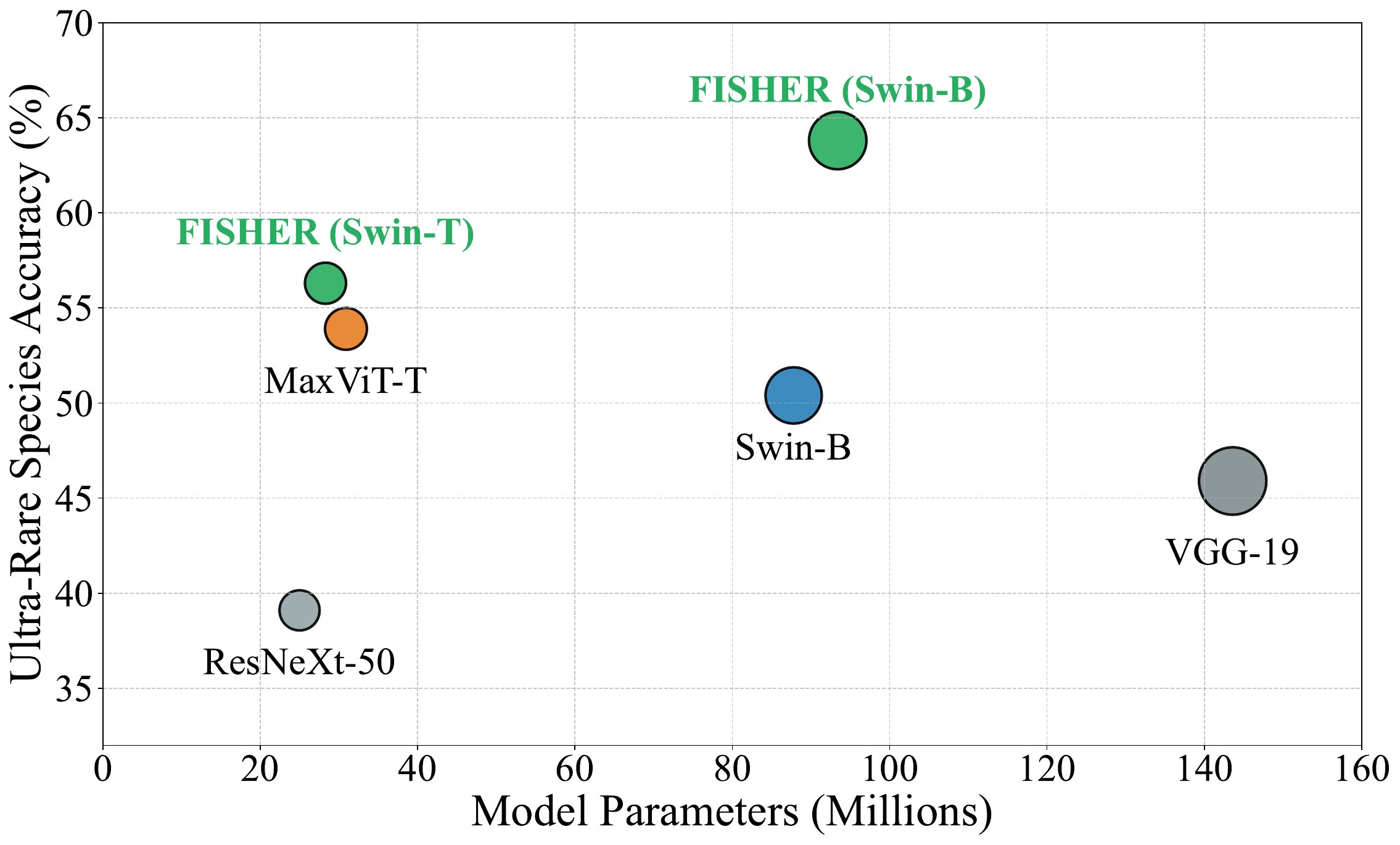}
    \caption{Ultra-rare species accuracy versus model complexity.}
    \label{fig:ultrarare_complexity}
    \vspace{-1em}
\end{figure}

%------------------------ Table XV -------------------------
\begin{table}[]
\centering
\textcolor{black}{
\caption{Efficiency comparison at Fish-Vista scale on a single GPU NVIDIA RTX~4070}
\label{tab:efficiency}
\renewcommand{\arraystretch}{1.15}
\resizebox{\columnwidth}{!}{%
\begin{tabular}{l c c c c}
\toprule
Model & Params (M) & GFLOPs & Latency$_{b=1}$ (ms) & Throughput$_{b=32}$ (img/s) \\
\midrule
RegNetY-4G            & 24.3  & 8.0  & 3.08 & 1083.0 \\
ResNeXt-50            & 31.8  & 8.6  & 1.53 & 855.8 \\
MaxViT-T              & 32.6  & 10.7 & 6.61 & 442.1 \\
ViT-B/16              & 89.1  & 33.7 & 4.16 & 384.2 \\
Swin-B (single-task)  & 91.2  & 30.4 & 5.15 & 304.6 \\
VGG-19                & 157.3 & 39.3 & 5.63 & 447.2 \\
\midrule
Parallel MTL (Connected)$^\dagger$ & 96.8 & 39.8 & 6.28 & 261.9 \\
$\FISHER$ (Detached)$^\dagger$     & 96.8 & 39.8 & 6.28 & 261.9 \\
\bottomrule
\end{tabular}%
}
}
\vspace{-1em}
\end{table}

\subsection{Model Complexity Analysis}
\label{subsec:complexity_ultrarare}

We analyze model complexity via ultra-rare species performance in Fig.~\ref{fig:ultrarare_complexity}, plotting accuracy against model size. Conventional CNN and Transformer baselines show saturation, where larger models do not consistently improve performance on rare classes. In contrast, $\FISHER$ achieves markedly higher ultra-rare accuracy with comparable model size. Notably, the Swin-B variant attains the best performance despite having fewer parameters than larger models such as VGG-19. These results indicate that the gains stem from the proposed hierarchical multi-task design rather than increased model capacity. \ph{Table~\ref{tab:efficiency} reports parameters, FLOPs, and inference latency: relative to the plain Swin-B backbone, the three-head hierarchy adds only $+5.6$M parameters ($+6\%$) and $+1.13$ ms latency, and gradient decoupling is \emph{FLOP-free}, as it prunes only backward-pass edges, so $\FISHER$ and the parallel MTL variant are computationally identical at inference.}

\section{Conclusion}
\label{sec:conclusion}
In this work, we have addressed the key challenge of fine-grained aquatic species classification by mitigating destructive gradient interference in hierarchical multi-task learning. We have identified gradient conflict as a primary factor limiting performance, where strong gradients from the classification task disrupted the representations required for segmentation and trait identification.
To address this issue, we have introduced $\FISHER$, a hierarchical multi-task framework that enforces a unidirectional information flow from segmentation to traits and then to species classification. Strategic gradient detachment was applied at task boundaries to isolate lower-level representations from higher-level classification gradients, thereby reducing negative transfer. In addition, we have incorporated learnable prototypes with orthogonality constraints to represent anatomical parts in the semantic segmentation space, and employed homoscedastic uncertainty weighting to dynamically balance task contributions during training.
Extensive experiments on the large-scale Fish-Vista dataset demonstrated that the proposed framework consistently outperformed state-of-the-art baselines in both multi-task and single-task settings. These results highlighted the effectiveness of hierarchical task structuring for learning biologically meaningful representations in long-tailed aquatic datasets.

\bibliographystyle{IEEEtran}
\bibliography{references}

\end{document}